\documentclass[]{jair}

\setcopyright{cc}
\acmDOI{10.1613/jair.1.20870}

\JAIRAE{Daniel Harabor}
\JAIRTrack{Multi-Agent Path Finding} %
\acmVolume{86}
\acmArticle{56}
\acmMonth{08}
\acmYear{2026}

\RequirePackage[
  datamodel=acmdatamodel,
  style=acmauthoryear,
  backend=biber,
  giveninits=true,
  uniquename=init
  ]{biblatex}

\usepackage{fontawesome5}

\usepackage{latex-common/ps}

\hypersetup{
    pdftitle={Simultaneous Computation with Multiple Prioritizations in Multi-Agent Motion Planning},
    pdfauthor={Patrick Scheffe},
}

\DeclareAcronym{admm}{
    short = ADMM,
    long  = Alternating Direction Method of Multipliers
}
\DeclareAcronym{cav}{
    short = CAV,
    long  = connected and automated vehicle,
}
\DeclareAcronym{cbs}{
    short = CBS,
    long  = conflict-based search,
}
\DeclareAcronym{cg}{
    short = CG,
    long  = center of gravity
}
\DeclareAcronym{cdmpc}{
    short = coop. DMPC,
    long  = cooperative distributed \acl{mpc}
}
\DeclareAcronym{cmpc}{
    short = CMPC,
    long  = centralized \acl{mpc}
}
\DeclareAcronym{cpm}{
    short = CPM,
    long  = Cyber-Physical Mobility
}
\DeclareAcronym{cpmlab}{
    short = CPM Lab,
    long  = \acl{cpm} Lab
}
\DeclareAcronym{ctg}{
    short = CTG,
    long  = cost to go
}
\DeclareAcronym{ctc}{
    short = CTC,
    long  = cost to come
}
\DeclareAcronym{dag}{
    short = DAG,
    long  = directed acyclic graph
}
\DeclareAcronym{dds}{
    short = DDS,
    long  = data distribution service
}
\DeclareAcronym{dmpc}{
    short = DMPC,
    long  = distributed \acl{mpc}
}
\DeclareAcronym{dpc}{
    short = DPC,
    long  = distributed predictive control
}
\DeclareAcronym{fsa}{
    short = FSA,
    long  = finite state automaton,
    short-indefinite = an,
}
\DeclareAcronym{fca}{
    short = FCA,
    long = future collision assessment,
    short-indefinite = an,
}
\DeclareAcronym{ffo}{
    short = FFO,
    long  = first-fit ordering,
    short-indefinite = an,
}
\DeclareAcronym{fov}{
    short = FOV,
    long  = field of view,
    short-indefinite = an,
}
\DeclareAcronym{fpv}{
    short = FPV,
    long  = first-person view,
    short-indefinite = an,
}
\DeclareAcronym{hdv}{
    short = HDV,
    long  = human-driven vehicle,
    short-indefinite = an,
}
\DeclareAcronym{hil}{
    short = HiL,
    long  = hardware-in-the-loop,
}
\DeclareAcronym{hlc}{
    short = HLC,
    long  = high-level controller,
    short-indefinite = an,
}
\DeclareAcronym{ldo}{
    short = LDO,
    long  = largest degree ordering,
    short-indefinite = an,
}
\DeclareAcronym{llc}{
    short = LLC,
    long  = low-level controller,
    short-indefinite = an,
}
\DeclareAcronym{lwa}{
    short = LWA*,
    long  = lazy weighted A*,
    short-indefinite = an,
}
\DeclareAcronym{lsp}{
    short = LazySP,
    long  = lazy shortest path,
    short-indefinite = an,
}
\DeclareAcronym{lra}{
    short = LRA*,
    long  = lazy receding horizon A*,
    short-indefinite = an,
}
\DeclareAcronym{mamp}{
    short = MAMP,
    long  = multi-agent motion planning,
    short-indefinite = a,
}
\DeclareAcronym{mapf}{
    short = MAPF,
    long  = multi-agent path finding,
    short-indefinite = a,
}
\DeclareAcronym{mas}{
    short = MAS,
    long  = multi-agent system,
    short-indefinite = an,
}
\DeclareAcronym{mcts}{
    short = MCTS,
    long  = Monte Carlo tree search,
    short-indefinite = an,
}
\DeclareAcronym{mip}{
    short = MIP,
    long  = mixed integer programming,
    short-indefinite = an,
}
\DeclareAcronym{mil}{
    short = MiL,
    long  = model-in-the-loop,
}
\DeclareAcronym{milp}{
    short = MILP,
    long  = mixed integer linear programming,
    short-indefinite = an,
}
\DeclareAcronym{mld}{
    short = MLD,
    long  = mixed logical dynamical,
    short-indefinite = an,
}
\DeclareAcronym{mlc}{
    short = MLC,
    long  = mid-level controller,
    short-indefinite = an,
}
\DeclareAcronym{mp}{
    short = MP,
    long  = motion primitive,
    short-indefinite = an,
}
\DeclareAcronym{mpa}{
    short = MPA,
    long  = motion primitive automaton,
    short-indefinite = an,
}
\DeclareAcronym{mpc}{
    short = MPC,
    long  = model predictive control,
    short-indefinite = an,
}
\DeclareAcronym{ncs}{
    short = NCS,
    long  = networked control system,
    short-indefinite = an,
}
\DeclareAcronym{nlp}{
    short = NLP,
    long  = nonlinear programming,
    short-indefinite = an,
}
\DeclareAcronym{ocp}{
    short = OCP,
    long  = optimal control problem,
    short-indefinite = an,
    long-indefinite = an,
}
\DeclareAcronym{odd}{
    short = ODD,
    long  = operational design domain,
    short-indefinite = an,
    long-indefinite = an,
}
\DeclareAcronym{ode}{
    short = ODE,
    long  = ordinary differential equation,
    short-indefinite = an,
    long-indefinite = an,
}
\DeclareAcronym{pdmpc}{
    short = \mbox{P-DMPC},
    long  = prioritized \acl{dmpc}
}
\DeclareAcronym{pil}{
    short = PiL,
    long  = processor-in-the-loop
}
\DeclareAcronym{pp}{
    short = PP,
    long  = prioritized planning
}
\DeclareAcronym{qp}{
    short = QP,
    long  = quadratic programming,
}
\DeclareAcronym{rhgs}{
    short = RHGS,
    long  = receding horizon graph search,
    short-indefinite = an,
}
\DeclareAcronym{rhc}{
    short = RHC,
    long  = receding horizon control,
    short-indefinite = an,
}
\DeclareAcronym{rrt}{
    short = RRT,
    long  = rapidly-exploring random tree,
    short-indefinite = an,
}
\DeclareAcronym{rss}{
    short = RSS,
    long  = responsibility-sensitive safety,
    short-indefinite = an,
}
\DeclareAcronym{rti}{
    short = RTI,
    long  = real-time iteration,
    short-indefinite = an,
}
\DeclareAcronym{scdmpc}{
    short = SC-DMPC,
    long = Synchronization-Based Cooperative Distributed Model Predictive Control,
    short-indefinite = an
}
\DeclareAcronym{scp}{
    short = SCP,
    long  = sequential convex programming,
    short-indefinite = an,
}
\DeclareAcronym{scr}{
    short = SCR,
    long  = sequential convex restriction,
    short-indefinite = an,
}
\DeclareAcronym{sdo}{
    short = SDO,
    long  = saturation degree ordering,
    short-indefinite = an,
}
\DeclareAcronym{sgs}{
    short = SGS,
    long  = state-of-the-art graph search,
    short-indefinite = an,
}
\DeclareAcronym{sil}{
    short = SiL,
    long  = software-in-the-loop,
}
\DeclareAcronym{sl}{
    short = SL,
    long  = sequential linearization,
    short-indefinite = an,
}
\DeclareAcronym{sqp}{
    short = SQP,
    long  = sequential quadratic programming,
    short-indefinite = an,
}
\DeclareAcronym{tsp}{
    short = TSP,
    long  = traveling salesman problem,
}
\DeclareAcronym{uav}{
    short = UAV,
    long  = unmanned aerial vehicle,
    long-indefinite = an,
}
\DeclareAcronym{udlab}{
    short = IDS3C,
    long  = Information and Decision Science Scaled Smart City,
}
\DeclareAcronym{xil}{
    short = XiL,
    long  = X-in-the-loop,
    long-indefinite = an,
}

\newcommand{\anAgent}{\ensuremath{i}}
\newcommand{\anotherAgent}{\ensuremath{j}}
\newcommand{\timestep}{\ensuremath{k}}
\newcommand{\timestepIterator}{\ensuremath{l}}

\newcommand{\ofAgent}[1]{\ensuremath{^{(#1)}}}
\newcommand{\ofAgentInAgent}[2]{\ensuremath{^{(#1 \,\vert\, #2)}}}
\newcommand{\forTimeAtTime}[2]{\ensuremath{_{#1 \,\vert\, #2}}}

\newcommand{\seqCoupling}{\ensuremath{\leftarrow}}

\newglossaryentry{matrix:Adjacency}{
	name=\ensuremath{\bm{D}},
	description={Adjacency matrix},
	sort={D},
    type=symbol
}
\newcommand{\matAdjacency}{\gls{matrix:Adjacency}}
\newcommand{\matAdjacencyElement}[1]{\glslink{matrix:Adjacency}{\matAdjacency_{#1}}}

\newglossaryentry{set:realNumbers}{
	name=\ensuremath{\mathbb{R}},
	description={Set of real numbers},
	sort={real numbers},
    type=symbol
}
\newcommand{\setRealNumbers}{\gls{set:realNumbers}}

\newglossaryentry{set:naturalNumbers}{
	name=\ensuremath{\mathbb{N}},
	description={Set of natural numbers},
	sort={natural numbers},
    type=symbol
}
\newcommand{\setNaturalNumbers}{\gls{set:naturalNumbers}}

\newglossaryentry{set:systemStates}{
	name=\ensuremath{\mathcal{S}},
	description={Set of system states},
	sort={set of system states},
    type=symbol
}

\newglossaryentry{set:bigO}{
	name=\ensuremath{O},
	description={Big O},
	sort={O},
    type=symbol
}
\newcommand{\bigO}{\gls{set:bigO}}

\newglossaryentry{scalar:Weight}{
	name=\ensuremath{w},
	description={Weight},
	sort={weight},
    type=symbol
}

\newglossaryentry{scalar:NumberOfAgents}{
    name=\ensuremath{N_A},
    description={Number of agents},
    sort={Number of agents},
    type=symbol
}
\newcommand{\numAgents}{\gls{scalar:NumberOfAgents}}

\newglossaryentry{graph:path}{
    name=\ensuremath{\pi},
    description={Path},
    sort={Path},
    type=symbol
}
\newcommand{\graphPath}{\gls{graph:path}}

\newglossaryentry{scalar:NumberOfVerticesInPath}{
    name=\ensuremath{N_{\graphPath}},
    description={Number of vertices in path $\graphPath$, or length},
    sort={Number of vertices in path},
    type=symbol
}

\newglossaryentry{sym:horizonControl}{
	name=\ensuremath{N_u},
	description={Control horizon in model predictive control},
	sort={Nu},
    type=symbol
}
\newcommand{\horizonControl}{\gls{sym:horizonControl}}

\newglossaryentry{sym:horizonPrediction}{
	name=\ensuremath{N_p},
	description={Prediction horizon in model predictive control},
	sort={Np},
    type=symbol
}
\newcommand{\horizonPrediction}{\gls{sym:horizonPrediction}}

\newglossaryentry{sym:vehicleOrientation}{
	name=\ensuremath{\psi},
	description={Vehicle orientation},
	sort={psi},
    type=symbol
}

\newglossaryentry{sym:sysModelContinuous}{
    name=\ensuremath{f},
    description={Continuous-time system model},
    sort={f continuous-time},
    type=symbol
}
\newcommand{\sysModelContinuous}{\gls{sym:sysModelContinuous}}

\newglossaryentry{sym:sysModelDiscrete}{
    name=\ensuremath{f_{d}},
    description={Discrete-time system model},
    sort={f discrete-time},
    type=symbol
}
\newcommand{\sysModelDiscrete}{\gls{sym:sysModelDiscrete}}

\newglossaryentry{sym:sysControlInputs}{
	name=\ensuremath{\bm{u}},
	description={System control inputs},
	sort=u,
    type=symbol
}
\NewDocumentCommand{\sysControlInputs}{ o }{\glslink{sym:sysControlInputs}{%
    \IfNoValueTF{#1}%
        {\ensuremath{\bm{u}}}%
        {\ensuremath{\bm{u}\ofAgent{#1}}}%
}}

\newglossaryentry{sym:outputs}{
	name=\ensuremath{\bm{y}},
	description={System outputs},
	sort={y},
    type=symbol
}

\newglossaryentry{sym:sysSpeed}{
	name=\ensuremath{\mathrm{v}},
	description={Vehicle speed},
	sort={v},
    type=symbol
}
\newcommand{\sysSpeed}{\gls{sym:sysSpeed}}

\newglossaryentry{sym:inSpeed}{
	name=\ensuremath{u_{\sysSpeed}},
	description={Vehicle input speed},
	sort={uv},
    type=symbol
}
\newcommand{\sysInSpeed}{\gls{sym:inSpeed}}

\newglossaryentry{sym:steering-angle}{
	name=\ensuremath{\delta},
	description={Vehicle steering angle},
	sort={delta},
    type=symbol
}

\newglossaryentry{sym:inSteering}{
	name=\ensuremath{u_{\delta}},
	description={Vehicle input steering angle},
	sort={ud},
    type=symbol
}
\newcommand{\sysInSteering}{\gls{sym:inSteering}}

\newglossaryentry{sym:nColors}{
	name=\ensuremath{N_c},
	description={Number of colors},
	sort={Number of colors},
    type=symbol
}

\newglossaryentry{sym:nStates}{
	name=\ensuremath{n},
	description={Number of states of a dynamical system},
	sort={Number of states},
    type=symbol
}
\newcommand{\numStates}{\gls{sym:nStates}}

\newglossaryentry{sym:nInputs}{
    name=\ensuremath{m},
    description={Number of inputs of a dynamical system},
    sort={m number of inputs},
    type=symbol
}
\newcommand{\numInputs}{\gls{sym:nInputs}}

\newglossaryentry{sym:nLevels}{
	name=\ensuremath{N_{c}},
	description={Number of computation levels},
	sort={Number of computation levels},
    type=symbol
}
\newcommand{\numLevels}{\gls{sym:nLevels}}

\newglossaryentry{sym:nLevelsAllowed}{
	name=\ensuremath{N_{\text{CL,al.}}},
	description={Allowed number of computation levels},
	sort={Number of computation levels allowed},
    type=symbol
}

\newglossaryentry{sym:numGroups}{
	name=\ensuremath{N_{g}},
	description={Number of parallelly computing groups of agents},
	sort={Number of groups},
    type=symbol
}

\newglossaryentry{sym:fcnPrio}{
    name=\ensuremath{p},
    description={Prioritization function},
    sort={Prioritization function},
    type=symbol
}
\newcommand{\fcnPrio}{\gls{sym:fcnPrio}}

\newglossaryentry{sym:tSample}{
	name=\ensuremath{T_s},
	description={Sample Time},
	sort={T sample},
    type=symbol
}
\newcommand{\tSample}{\gls{sym:tSample}}

\newglossaryentry{sym:tSolve}{
	name=\ensuremath{T_\text{sol.}},
	description={Computation time \tSolveB{\anAgent} that agent $\anAgent$ needs to solve its \ac{ocp}},
	sort={T solve},
    type=symbol
}

\newcommand{\tSolveB}[1]{\glslink{sym:tSolve}{\ensuremath{\ensuremath{T_\text{sol.}}^{(#1)}}}}

\newglossaryentry{sym:tSolveUpper}{
	name=\ensuremath{T_\text{sol.,max}},
	description={Upper computation time $T_\text{sol.,max}\ofAgent{\anAgent}$ that agent $\anAgent$ needs to solve it \ac{ocp}},
	sort={T solve upper},
    type=symbol
}

\newglossaryentry{sym:vertices}{
	name=\ensuremath{\mathcal{V}},
	description={Set of vertices},
	sort={Vertices},
    type=symbol
}
\newcommand{\setVertices}{\gls{sym:vertices}}
\newcommand{\setAgents}{\setVertices}

\newglossaryentry{sym:predecessors}{
	name=\ensuremath{\setVertices\ofAgent{i\leftarrow}},
	description={Set of predecessors of vertex $i$},
	sort={Vertices 1},
    type=symbol
}
\NewDocumentCommand{\setPredecessors}{ mo }{\glslink{sym:predecessors}{%
    \IfNoValueTF{#2}%
        { \ensuremath{\setVertices\ofAgent{#1\leftarrow}} }%
        { \ensuremath{\setVertices\ofAgent{#1\leftarrow}_{#2}} }%
}}

\newcommand{\helpSetPredecessorsPar}[1]{\ensuremath{\setVertices^{(#1\leftarrow)}_{\text{par.}}}}
\newglossaryentry{sym:predecessorsPar}{
	name=\ensuremath{\helpSetPredecessorsPar{i}},
	description={Set of predecessors of vertex $i$ that have parallel couplings with it},
	sort={Vertices 2},
    type=symbol
}

\newcommand{\helpSetPredecessorsSeq}[1]{\ensuremath{\setVertices^{(#1\leftarrow)}_{\text{seq.}}}}
\newglossaryentry{sym:predecessorsSeq}{
	name=\ensuremath{\helpSetPredecessorsSeq{i}},
	description={Set of predecessors of vertex $i$ that have sequential couplings with it},
	sort={Vertices 3},
    type=symbol
}

\newglossaryentry{sym:successors}{
	name=\ensuremath{\setVertices\ofAgent{i\rightarrow}},
	description={Set of successors of vertex $i$},
	sort={Vertices 4},
    type=symbol
}
\NewDocumentCommand{\setSuccessors}{ mo }{\glslink{sym:successors}{%
    \IfNoValueTF{#2}%
        { \ensuremath{\setVertices\ofAgent{#1\rightarrow}} }%
        { \ensuremath{\setVertices\ofAgent{#1\rightarrow}_{#2}} }%
}}

\newglossaryentry{sym:neighbors}{
	name=\ensuremath{\setVertices\ofAgent{i}},
	description={Set of neighbors of vertex $i$},
	sort={Vertices 0},
    type=symbol
}
\newcommand{\setNeighbors}[1]{\glslink{sym:neighbors}{\ensuremath{\setVertices\ofAgent{#1}}}}

\newglossaryentry{sym:degree}{
	name=\ensuremath{d^{(i)}},
	description={Degree of vertex $i$. Sum of in-degree and out-degree},
	sort=degree,
    type=symbol
}
\newcommand{\vertexDegree}[1]{\glslink{sym:degree}{\ensuremath{d^{(#1)}}}}

\newcommand{\helpVertexInDegree}[1]{\ensuremath{d^{(#1\leftarrow)}}}
\newglossaryentry{sym:inDegree}{
    name=\helpVertexInDegree{i},
    description={In-degree of vertex $i$},
    sort={degree in},
    type=symbol
}
\newcommand{\vertexInDegree}[1]{\glslink{sym:inDegree}{\helpVertexInDegree{#1}}}

\newcommand{\helpVertexOutDegree}[1]{\ensuremath{d^{(#1\rightarrow)}}}
\newglossaryentry{sym:outDegree}{
    name=\helpVertexOutDegree{i},
    description={Out-degree of vertex $i$},
    sort={degree out},
    type=symbol,
}
\newcommand{\vertexOutDegree}[1]{\glslink{sym:outDegree}{\helpVertexOutDegree{#1}}}

\newglossaryentry{sym:matLevels}{
	name=\ensuremath{\bm{L}},
	description={Matrix of computation levels},
	sort=L,
    type=symbol
}

\newglossaryentry{sym:tComp}{
	name=\ensuremath{T},
	description={Computation time},
	sort={T},
    type=symbol
}
\newcommand{\tComp}{\gls{sym:tComp}}

\newglossaryentry{sym:tCompNcs}{
	name=\ensuremath{T},
	description={Computation time of \iac{ncs}},
	sort={T},
    type=symbol
}
\newcommand{\tCompNcs}{\gls{sym:tCompNcs}}

\newglossaryentry{graph:Undirected}{
	name=\ensuremath{\mathcal{G}},
	description={Undirected Graph},
	sort={graph1},
    type=symbol
}
\newcommand{\graphUndirected}{\gls{graph:Undirected}}

\newglossaryentry{graph:Directed}{
    name=\ensuremath{\vec{\gls*{graph:Undirected}}},
	description={Directed Graph},
	sort={graph2},
    type=symbol
}
\newcommand{\graphDirected}{\gls{graph:Directed}}

\newglossaryentry{mat:edgeUtilities}{
    name=\ensuremath{M_\text{u}},
	description={Edge utility matrix},
	sort={matrix edge utilities},
    type=symbol
}

\newglossaryentry{sym:setColors}{
    name=\ensuremath{\mathcal{C}},
    description={Set of colors},
    sort=Colors,
    type=symbol
}

\newglossaryentry{sym:varControlInvariantSet}{
	name=\ensuremath{\mathcal{C}_{\text{inv}}},
	description={Control invariant set},
	sort={Control invariant set},
    type=symbol
}

\newglossaryentry{set:Weights}{
	name=\ensuremath{\mathcal{W}},
	description={Set of weights in a weighted graph},
    sort={Weights},
    type=symbol
}

\newglossaryentry{set:Edges}{
	name=\ensuremath{\mathcal{E}},
	description={Set of edges; used to indicate that only undirected edges exist},
    sort={Edges},
    type=symbol
}
\newcommand{\setEdges}{\gls{set:Edges}}

\newglossaryentry{sym:varEdgeUndirected}{
	name=\ensuremath{(i - j)},
	description={Undirected between vertex $i$ and vertex $j$},
	sort={edge1},
    type=symbol
}
\newcommand{\edgeUndirected}[2]{\glslink{sym:varEdgeUndirected}{\ensuremath{(#1 - #2)}}}

\newglossaryentry{sym:setEdgesDirected}{
	name=\ensuremath{\vec{\gls*{set:Edges}}},
	description={Set of directed edges},
	sort={Edges directed},
    type=symbol
}
\newcommand{\setEdgesDirected}{\gls{sym:setEdgesDirected}}

\newglossaryentry{sym:varEdge}{
	name=\ensuremath{(i \rightarrow j)},
	description={Directed edge from vertex $i$ to vertex $j$},
	sort={edge2},
    type=symbol
}
\newcommand{\edgeDirected}[2]{\glslink{sym:varEdge}{\ensuremath{(#1 \rightarrow #2)}}}

\newglossaryentry{sym:fnReorder}{
	name=\ensuremath{f_r},
	description={Reordering function for graph color values},
	sort=fr,
    type=symbol
}

\newglossaryentry{sym:fcnObjective}{
    name=\ensuremath{J},
    description={Objective function of an optimization problem},
    sort=J,
    type=symbol
}

\NewDocumentCommand{\fcnObjective}{ oo }{\glslink{sym:fcnObjective}{%
    \IfNoValueTF{#1}%
    {\ensuremath{J}}%
        {\IfNoValueTF{#2}%
            { \ensuremath{J\ofAgent{#1}} }%
            { \ensuremath{J\ofAgent{#1}_{#2}} }%
        }
}}
\NewDocumentCommand{\fcnObjectiveNcs}{ o }{\glslink{sym:fcnObjective}{%
    \IfNoValueTF{#1}%
    {\ensuremath{J}}%
        { \ensuremath{J_{#1}} }%
}}

\newglossaryentry{sym:fcnObjectiveState}{
    name=\ensuremath{\ell_{x}},
    description={Reference tracking objective function},
    sort={lx Reference tracking objective function},
    type=symbol
}
\NewDocumentCommand{\fcnObjectiveState}{ o }{\glslink{sym:fcnObjectiveState}{%
    \IfNoValueTF{#1}%
        {\ensuremath{\ell_{x}}}%
        {\ensuremath{\ell_{x}^{(#1)}}}%
}}

\newglossaryentry{sym:fcnObjectiveStateTerminal}{
    name=\ensuremath{\ell_{x,f}},
    description={Reference tracking objective terminal function},
    sort={lf Reference tracking objective terminal function},
    type=symbol
}
\NewDocumentCommand{\fcnObjectiveStateTerminal}{ o }{\glslink{sym:fcnObjectiveStateTerminal}{%
    \IfNoValueTF{#1}%
        {\ensuremath{\ell_{x,f}}}%
        {\ensuremath{\ell_{x,f}^{(#1)}}}%
}}

\newglossaryentry{sym:fcnObjectiveInput}{
    name=\ensuremath{\ell_{u}},
    description={Input change objective function},
    sort={lu Input change objective function},
    type=symbol
}
\NewDocumentCommand{\fcnObjectiveInput}{ o }{\glslink{sym:fcnObjectiveInput}{%
    \IfNoValueTF{#1}%
        {\ensuremath{\ell_{u}}}%
        {\ensuremath{\ell_{u}^{(#1)}}}%
}}

\newglossaryentry{sym:fcnObjectiveCoupling}{
    name=\ensuremath{\ell_\text{c}},
    description={Coupling objective function},
    sort={lc Coupling objective function},
    type=symbol
}
\NewDocumentCommand{\fcnObjectiveCoupling}{ oo }{\glslink{sym:fcnObjectiveCoupling}{%
    \IfNoValueTF{#1}%
        {\ensuremath{\ell_\text{c}}}%
        {\ensuremath{\ell_\text{c}^{(#1,#2)}}}%
}}

\newglossaryentry{sym:fcnConstraintCoupling}{
    name=\ensuremath{c_\text{c}},
    description={Coupling constraint function},
    sort={cc Coupling constraint function},
    type=symbol
}
\NewDocumentCommand{\fcnConstraintCoupling}{ oo }{\glslink{sym:fcnConstraintCoupling}{%
    \IfNoValueTF{#1}%
        {\ensuremath{c_\text{c}}}%
        {\ensuremath{c_\text{c}^{(#1,#2)}}}%
}}

\newglossaryentry{sym:prediction}{
	name=\ensuremath{\bm{x}\ofAgentInAgent{j}{i}\forTimeAtTime{\cdot}{k}},
	description={Prediction of agent $j$ in agent $i$ at time $k$},
	sort=x,
    type=symbol,
}
\NewDocumentCommand{\agentPrediction}{ oo }{\glslink{sym:prediction}{%
    \IfNoValueTF{#1}%
        {\ensuremath{ \bm{x}\forTimeAtTime{\cdot}{\timestep} }}%
        {\IfNoValueTF{#2}%
            { \ensuremath{ \bm{x}\forTimeAtTime{\cdot}{\timestep}\ofAgent{#1} } }%
            { \ensuremath{ \bm{x}\forTimeAtTime{\cdot}{\timestep}\ofAgentInAgent{#1}{#2} } }%
        }
}}

\newglossaryentry{sym:state}{
	name=\ensuremath{\bm{x}},
	description={System state},
	sort=x,
    type=symbol
}
\newcommand{\sysState}{\gls{sym:state}}

\newglossaryentry{sym:stateAgent}{
	name=\ensuremath{\sysState^{(i)}_{(k)}},
	description={System state of agent $i$ at time $k$},
	sort=x,
    type=symbol,
}

\newglossaryentry{sym:ref}{
	name=\ensuremath{\bm{r}^{(i)}_{k}},
	description={System state reference of agent $i$ at time $k$},
	sort=x ref,
    type=symbol
}
\newcommand{\sysRef}{ \glslink{sym:ref}{\ensuremath{\bm{r}}} }

\newglossaryentry{sym:setReachable}{
	name=\ensuremath{\mathcal{R}^{(i)}},
	description={reachable set of agent $i$},
	sort={Reachable set},
    type=symbol
}
\newcommand{\setReachable}{\glslink{sym:setReachable}{\ensuremath{\mathcal{R}}}}
\newcommand{\setReachableB}[1]{\glslink{sym:setReachable}{\ensuremath{\mathcal{R}^{(#1)}}}}

\newglossaryentry{set:occupiedArea}{
	name=\ensuremath{\mathcal{O}^{(i)}},
	description={Set of the occupied area of the predicted trajectory of agent $\anAgent$},
	sort={occupied area},
    type=symbol
}

\newglossaryentry{set:feasibleStates}{
	name=\ensuremath{\mathcal{X}},
	description={set of feasible states},
	sort={x},
    type=symbol
}
\newcommand{\setFeasibleStates}{\gls{set:feasibleStates}}

\newglossaryentry{set:feasibleStatesTerminal}{
	name=\ensuremath{\mathcal{X}_f},
	description={set of feasible states at the prediction horizon},
	sort={x},
    type=symbol
}
\newcommand{\setFeasibleStatesTerminal}{\gls{set:feasibleStatesTerminal}}

\newglossaryentry{set:feasibleInputs}{
	name=\ensuremath{\mathcal{U}},
	description={set of feasible inputs},
	sort={u},
    type=symbol
}
\newcommand{\setFeasibleInputs}{\gls{set:feasibleInputs}}

\newglossaryentry{sym:numStatesConfSpace}{
    name=\ensuremath{n_p},
    description={Number of states that are in the conflictual space},
    sort={n number of states that are in the conflictual space},
    type=symbol
}

\newglossaryentry{sym:fnProj}{
    name=\text{proj},
    description={A function that projects a reachable set of system states in the conflictual space},
    sort={Project function},
    type=symbol
}

\newglossaryentry{sym:setClasses}{
	name=\ensuremath{\mathfrak{V}},
	description={Set of classes of parallelizable agents},
	sort={set class agents},
    type=symbol
}
\newcommand{\setClasses}{\gls{sym:setClasses}}

\newglossaryentry{sym:classAgents}{
	name=\ensuremath{\setVertices_{c}},
	description={Class of parallelizable agents},
	sort={class agents},
    type=symbol
}
\NewDocumentCommand{\classAgents}{ o }{\glslink{sym:classAgents}{%
    \IfNoValueTF{#1}%
        {\ensuremath{\setVertices_{c}}}%
        {\ensuremath{\setVertices_{#1}}}%
}}

\newglossaryentry{sym:sequenceClasses}{
	name=\ensuremath{(s_z)_{z=1}^{\numLevels}},
	description={Sequence of classes of parallelizable agents},
	sort={sequence class agents},
    type=symbol
}
\newcommand{\sequenceClasses}{\gls{sym:sequenceClasses}}
\newcommand{\sequenceElement}[1]{\glslink{sym:matSchedule}{s_{#1}}}

\newglossaryentry{sym:matSchedule}{
	name=\ensuremath{\bm{S}},
	description={Matrix of computation schedule},
	sort=S,
    type=symbol
}
\newcommand{\matSchedule}{\gls{sym:matSchedule}}
\newcommand{\matScheduleElement}[1]{\glslink{sym:matSchedule}{s_{#1}}}
\NewDocumentCommand{\matScheduleVector}{ mo }{\glslink{sym:matSchedule}{%
    \IfNoValueTF{#2}%
        { \bm{s}_{#1} }%
        { \bm{s}_{#1}^{#2} }%
}}

\newglossaryentry{sym:setSequences}{
	name=\ensuremath{\mathcal{S}},
	description={Set of sequences},
	sort={Set Sequences},
    type=symbol
}
\newcommand{\setSequences}{\gls{sym:setSequences}}

\newglossaryentry{sym:tSolveClass}{
	name=\ensuremath{T_\text{sol.}},
	description={Computation time \tSolveClass{\classAgents} that a agent class $\classAgents$ needs to solve their \acp{ocp}},
	sort={T solve class},
    type=symbol
}
\newcommand{\tSolveClass}[1]{\glslink{sym:tSolveClass}{\ensuremath{\ensuremath{T_\text{sol.}}^{[#1]}}}}

\newglossaryentry{sym:tCompNcsUpper}{
	name=\ensuremath{\hat{T}_{\text{NCS}}},
	description={Fixed computation time for \iac{ncs}},
	sort={T NCS fixed},
    type=symbol
}

\newglossaryentry{sym:latinSquare}{
	name=\ensuremath{\mathfrak{L}},
	description={Latin Square},
	sort={Latin Square},
    type=symbol
}

\newglossaryentry{def:agent}{
	name=agent,
	description={An agent is a system which is composed of at least one of the three elements: sensors, actuators, and a dynamic behavior.%
    },
}

\newglossaryentry{def:agentActive}{
	name=active agent,
	description={Active \glspl{def:agent} are \glspl{def:agent} which are connected using a communication
    network over which they can exchange data. The exchanged data is
    used by the \glspl{def:agent}' controllers to find appropriate inputs to achieve their
    goals while interacting with other \glspl{def:agent}.
    Additionally, active \glspl{def:agent} consider \glspl{def:agentPassive}},
    parent=def:agent,
}

\newglossaryentry{def:agentPassive}{
	name=passive agent,
	description={Passive \glspl{def:agent} are \glspl{def:agent} without networked control. However, they can communicate their data like current and future states to \glspl{def:agentActive}, or they can be detected by \glspl{def:agentActive}' sensors.%
    },
    parent=def:agent,
}

\newglossaryentry{def:distrutedSolutionQuality}{
	name=distributed solution quality,
	description={%
        The quality $q\in [0,1]$ of the solution in \ac{dmpc} is the networked objective function value ${\fcnObjective}_{c}$ for the solution of the corresponding \ac{cmpc} formulation divided by the objective function value ${\fcnObjective_d}$ for the solution of the \ac{dmpc} formulation
        \begin{equation}
            q = \frac{\fcnObjective_c}{\fcnObjective_d}.
        \end{equation}
    },
}

\newglossaryentry{def:mas}{
	name=multi-agent system,
	description={A system consisting of multiple \glspl{def:agent}.%
    },
}

\newglossaryentry{def:ncs}{
	name=networked control system,
	description={A system consisting of multiple \glspl{def:agentActive}.%
    },
}

\newglossaryentry{def:prediction}{
	name=prediction,
	description={
        A prediction $\agentPrediction\ofAgent{\anAgent}$ of \gls{def:agent} $\anAgent$ is its predicted state trajectory as obtained from the solution of its \ac{ocp} at time $\timestep$.
        A prediction $\agentPrediction\ofAgentInAgent{\anAgent}{\anotherAgent}$ of \gls{def:agent} $\anAgent$ for \gls{def:agent} $\anotherAgent$ is agent $\anotherAgent$'s state trajectory as viewed from agent $\anAgent$ at time $\timestep$. It is obtained by communication or by predicting \gls{def:agent} $\anotherAgent$'s state trajectory with its model using the solution to its \ac{ocp}.%
    },
}

\newglossaryentry{def:consistency}{
	name=prediction consistency,
	description={%
        A \ac{ncs} is prediction consistent at time step $\timestep$ if the \gls{def:prediction} $\agentPrediction\ofAgentInAgent{\anotherAgent}{\anAgent}$
            of all neighbors $\anotherAgent \in \setNeighbors{\anAgent}$
            in every agent $\anAgent\in\setAgents$
            equals their own \gls{def:prediction} $\agentPrediction\ofAgent{\anotherAgent}$, i.e.,
        \begin{equation}
            \agentPrediction\ofAgentInAgent{\anotherAgent}{\anAgent}=\agentPrediction\ofAgent{\anotherAgent}, \quad \forall \anAgent \in \setAgents, \forall \anotherAgent \in \setNeighbors{\anAgent}.
        \end{equation}%
    }
}

\newglossaryentry{def:ncsFeasible}{
	name=NCS-feasible,
	description={%
        The solutions $\sysControlInputs_{\cdot \vert \timestep}\ofAgent{\anAgent}$ of all agents $i\in\setAgents$ are \acs*{ncs}-feasible if the stacked solution vector $\bm{U}_{\cdot \vert \timestep} = \left( \sysControlInputs_{\cdot \vert \timestep}\ofAgent{1}, \ldots, \sysControlInputs_{\cdot \vert \timestep}\ofAgent{\numAgents} \right)\transposed$ satisfies all constraints of the corresponding central \acf*{ocp} considering all agents.%
    },
}

\newglossaryentry{def:feasibleAgent}{
	name=agent-feasible,
	description={%
        A solution is agent-feasible if it satisfies the constraints of to the corresponding agent's \ac{ocp}.%
    },
}

\newglossaryentry{def:networkedObjectiveFunction}{
	name=networked objective function,
	description={%
        The objective function value ${\fcnObjective}$ in \iac{ncs} formulation is the sum of all agent objective functions \fcnObjective[i]
        \begin{equation}
            \fcnObjective = \sum_{i}^{i\in\setAgents} \fcnObjective[i].
        \end{equation}
    },
}

\newglossaryentry{def:optimalPrioritization}{
	name=optimal prioritization,
	description={%
        The optimal prioritization results in the solution for every agent with the lowest networked objective function value obtainable by prioritization.%
    },
}

\newglossaryentry{def:graph_undirected}{
	name=undirected graph,
	description={%
        An undirected graph $\graphUndirected = \left(\setVertices,\setEdges\right)$ is a pair of two sets,
        the set of vertices $\setVertices=\set{1,\dots,\numAgents}$
        and the set of undirected edges $\setEdges \subseteq \setVertices \times \setVertices$.
        The edge between $i$ and $j$ is denoted by $\edgeUndirected{i}{j}$.
    },
}

\newglossaryentry{def:graph_directed}{
	name=directed graph,
	description={%
        A directed graph $\graphDirected = \left(\setVertices,\setEdgesDirected\right)$ is a pair of two sets,
        the set of vertices $\setVertices=\set{1,\dots,\numAgents}$
        and the set of directed edges $\setEdgesDirected \subseteq \setVertices \times \setVertices$.
        The edge from $i$ to $j$ is denoted by $\edgeDirected{i}{j}$.
        An oriented graph is a directed graph obtained from an undirected graph by replacing each edge $\edgeUndirected{i}{j}$ with either $\edgeDirected{i}{j}$ or $\edgeDirected{j}{i}$.
    },
}

\newglossaryentry{def:path}{
	name=path,
	description={%
        A path from vertex $\anAgent_1$ to vertex $\anAgent_{n+1}$ in a directed graph $\graphDirected$ is a sequence of $n \geq 1$ edges
        \begin{equation}
            (e_z)_{z=1}^{n}
            =
            \bigl(
                \edgeDirected{\anAgent_1}{\anAgent_2},
                \edgeDirected{\anAgent_2}{\anAgent_3},
                \ldots,
                \edgeDirected{\anAgent_{n}}{\anAgent_{n+1}}
            \bigr)
            ,
        \end{equation}
        with the domain $\set{1,\ldots,n}$ and the codomain $\setEdgesDirected$, connecting the pairwise distinct vertices $\anAgent_1, \ldots, \anAgent_{n+1}$.
        The length of the path is defined as the number of edges $n$.
    },
}

\newglossaryentry{def:graph:adjacency}{
	name=adjacency,
	description={%
    A vertex $j$ is a predecessor of vertex $i$ iff $\edgeDirected{j}{i}\in\setEdges$.
    The set of predecessors of vertex $i$ is denoted by
    \begin{equation}
        \setPredecessors{i}=\set{j \mid \edgeDirected{j}{i}\in\setEdges}.
    \end{equation}
    A vertex $j$ is a successor of vertex $i$ iff $\edgeDirected{i}{j}\in\setEdges$.
    The set of successors of vertex $i$ is denoted by
    \begin{equation}
        \setSuccessors{i}=\set{j \mid \edgeDirected{i}{j}\in\setEdges}.
    \end{equation}
    A vertex $j$ is a neighbor to or adjacent to vertex $i$ if it is either a predecessor or a successor.
    The set of neighbors of vertex $i$ is denoted by
    \begin{equation}
        \setNeighbors{i}= \setSuccessors{i} \cup \setPredecessors{i}.
    \end{equation}
    },
}

\newglossaryentry{def:graph:degree}{
	name=degree,
	description={%
        The degree
        \begin{equation}
            \vertexDegree{i} = \lvert \setNeighbors{i} \rvert
        \end{equation}
        of a vertex $i$ denotes the number of its adjacent vertices.
        The number of incoming edges called in-degree is denoted by
        \begin{equation}
            \vertexInDegree{i} = \lvert \setPredecessors{i} \rvert.
        \end{equation}
        The number of outgoing edges called out-degree is denoted by
        \begin{equation}
            \vertexOutDegree{i} = \lvert \setSuccessors{i} \rvert.
        \end{equation}
    },
}

\newglossaryentry{def:graph:diameter}{
	name=diameter,
	description={%
        The diameter of a graph is the greatest length of any shortest path between each pair of vertices $\setVertices \times \setVertices$ that is connected by a path.%
    },
}

\newglossaryentry{def:coupling_graph_undirected}{
	name=undirected coupling graph,
	description={%
        An undirected coupling graph
        $\graphUndirected(\timestep)=\bigl(\setVertices,\setEdges(\timestep)\bigr)$
        is a graph that represents the interaction between agents at time step $\timestep$.
        Vertices $\anAgent\in\setVertices=\set{1,\ldots,\numAgents}$ represent agents,
        and edges $\edgeUndirected{\anAgent}{\anotherAgent}\in\setEdges$ represent coupling objectives and constraints between agents.
    },
}

\newglossaryentry{def:coupling_graph_directed}{
	name=directed coupling graph,
	description={%
        A directed coupling graph
        $\graphDirected(\timestep)=\bigl(\setVertices,\setEdgesDirected(\timestep)\bigr)$
        is an oriented graph obtained from orienting the edges of an undirected coupling graph at time step $\timestep$.
        If an edge $\edgeDirected{\anAgent}{\anotherAgent}$ is directed from agent $\anAgent$ to agent $\anotherAgent$, then agent $\anotherAgent$ is responsible for considering the respective coupling objective and constraint in its planning problem.
    },
}

\newglossaryentry{def:matrix:Adjacency}{
	name=adjacency matrix,
	description={An adjacency matrix represents a graph with $\numAgents$ vertices in a matrix $\matAdjacency \in \set{0,1}^{\numAgents\times\numAgents}$ with entries
    \begin{equation}
        \matAdjacencyElement{ij} =
            \begin{cases}
                1 & \text{ if } \edgeDirected{i}{j} \in \setEdges \\
                0 & \text{ otherwise.}
            \end{cases}
    \end{equation}
    },
}

\newglossaryentry{def:tCompNcs}{
	name=computation time of \iac{ncs},
	description={%
        The computation time $\tCompNcs$ of \iac{ncs} is the time required for the \ac{ncs} to measure the states, formulate and solve the \ac{ocp}, apply the inputs to all agents, and communicate the required data. 
    },
}

\newglossaryentry{def:setReachable}{
	name=reachable set,
	description={%
        The reachable set of states $\setReachable$ of an agent from an initial time $t_{\text{init.}}$ to an end time $t_{\text{end}}$ is
            \begin{equation}\label{eq:setReachable}
                \setReachable_{[t_{\text{init.}},t_{\text{end}}] \mid t_{\text{init.}}} = \biggl\{ \int_{t_{\text{init.}}}^{t_{\text{end}}} \sysModelContinuous(\sysState,\sysControlInputs)dt
                \biggm| \sysState(t_{\text{init.}}) \in \setFeasibleStates(t_{\text{init.}}), \forall t: \sysControlInputs \in \setFeasibleInputs \biggr\},
            \end{equation}
        with the possible system initial states $\sysState(t_{\text{init.}})$ being bounded by its initially admissible set $\setFeasibleStates(t_{\text{init.}}) \subseteq \setRealNumbers^{\numStates}$, and the possible system control inputs $\sysControlInputs$ being bounded by its admissible set $\setFeasibleInputs \subseteq \setRealNumbers^{\numInputs}$.
    },
}

\newglossaryentry{def:conflictualDecisions}{
	name=conflictual decisions in \iac{ncs},
	description={%
        Consider two decisions made by two agents of \iac{ncs} at time step $\timestep$ with a duration $N_k$.
        They are deemed conflictual if the predicted outcome of the decisions violates the \ac{ncs}-feasibility at some point in time.%
    },
}

\newglossaryentry{def:conflictualSpace}{
	name=conflictual space of \iac{ncs},
	description={%
        In dynamic systems, the state space represents the set of all possible states the systems can occupy. 
        The conflictual space refers to a subset, or potentially the entirety, of this state space where whether decisions are conflictual is determined.
    },
}

\newglossaryentry{def:anytimePlanner}{
	name=Anytime trajectory planner,
	description={%
        An anytime trajectory planner is a trajectory planner that quickly identifies a feasible trajectory and incrementally improves it over time.
    },
}

\definecolor{color1}{RGB}{0, 84, 159}      %
\definecolor{color2}{RGB}{246, 168, 0}     %
\definecolor{color3}{RGB}{0, 152, 161}     %
\definecolor{color4}{RGB}{122, 111, 172}   %
\definecolor{color5}{RGB}{204, 7, 30}      %
\definecolor{color6}{RGB}{0, 97, 101}      %
\definecolor{color7}{RGB}{189, 205, 0}     %
\definecolor{color8}{RGB}{161, 16, 53}     %
\definecolor{color9}{RGB}{87, 171, 39}     %
\definecolor{color10}{RGB}{97, 33, 88}     %
\definecolor{color11}{RGB}{255, 237, 0}    %

\preprintfalse
\publishedfalse
\newcommand{\mydoi}{xxx}

\usepackage[T1]{fontenc}
\newcommand{\code}[1]{{\texttt{#1}}}

\newcommand{\videolink}{youtu.be/Mb59zQ3j3s0}

\glsdisablehyper %
\acsetup{make-links=false}

\newcommand{\Cpp}{C\raisebox{0.3ex}{\hspace{-0.1ex}\scriptsize\textbf{++}}}

\begin{document}

\title[Simultaneous Computation with Multiple Prioritizations]{Simultaneous Computation with Multiple Prioritizations in Multi-Agent Motion Planning}

\author{Patrick Scheffe}
\authornote{Corresponding Author. Also with Flanders Make@KU Leuven and RWTH Aachen University.}
\orcid{0000-0002-2707-198X}
\email{patrick.scheffe@kuleuven.be}
\affiliation{%
    \institution{KU Leuven}
    \city{Leuven}
    \country{Belgium}
}

\author{Julius Kahle}
\orcid{0000-0003-3986-1986}
\email{kahle@embedded.rwth-aachen.de}
\affiliation{%
    \institution{RWTH Aachen University}
    \city{Aachen}
    \country{Germany}
}

\author{Bassam Alrifaee}
\orcid{0000-0002-5982-021X}
\email{bassam.alrifaee@unibw.de}
\affiliation{%
    \institution{University of the Bundeswehr Munich}
    \city{Neubiberg}
    \country{Germany}
}

\renewcommand{\shortauthors}{Scheffe, Kahle \& Alrifaee}

\begin{abstract}
    \Ac{mapf} in large networks is computationally challenging.
    An approach for MAPF is \ac{pp}, in which agents plan sequentially according to their priority.
    Albeit a computationally efficient approach for MAPF, the solution quality strongly depends on the prioritization.
    Most prioritizations rely either on heuristics, which do not generalize well, or iterate to find adequate priorities, which costs computational effort.
    In this work, we show how agents can compute with multiple prioritizations simultaneously.
    Our approach is general as it does not rely on domain-specific knowledge.
    The context of this work is \ac{mamp} with a receding horizon subject to computation time constraints.
    MAMP considers the system dynamics in more detail compared to MAPF.
    In numerical experiments on MAMP, we demonstrate that our approach achieves near-optimal prioritization and outperforms state-of-the-art methods with only a minor increase in computation time.
    We show real-time capability in an experiment on a road network with ten vehicles in our Cyber-Physical Mobility Lab.
\end{abstract}

\acresetall

\received{27 October 2025}
\received[accepted]{02 August 2026}

\maketitle

\input{latex-common/copyright.tex}
\newcommand{\graphMap}      {\ensuremath{\graphUndirected_M}}
\newcommand{\setVerticesMap}{\ensuremath{\setVertices_M}}
\newcommand{\setEdgesMap}   {\ensuremath{\setEdges_M}}
\newcommand{\aVertex}       {\ensuremath{v}}
\newcommand{\anotherVertex} {\ensuremath{u}}

\section{Introduction}\label{sec:introduction}

\begin{figure}
    \centering
    \includegraphics{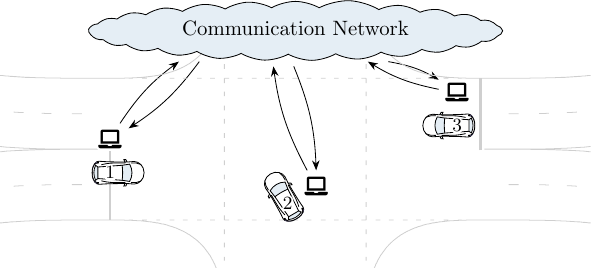}
    \caption{
        The application of this work is distributed \ac{mamp} for \acp{cav} without a central coordinator.
        Each agent knows only its own objective and computes only its own control input, but can communicate with other agents.
        Since the environment is dynamic and the planning horizon limited, agents plan frequently and should finish computation within a limited time duration.
    }
    \Description{
        Schematic top view of a road intersection with three vehicles, numbered one to three, approaching from different directions.
        Each vehicle carries a computer icon, indicating that it computes its own motion plan on board.
        A cloud labeled Communication Network is drawn above the road, and bidirectional arrows connect each vehicle's computer to the cloud, indicating that the vehicles exchange information with one another but not with any central coordinator.
    }
    \label{fig:application}
\end{figure}

\Ac{mapf} is a multi-agent planning problem in which agents must find a path from a start position to a target position on a discrete grid represented by a graph.
\Ac{mamp} for \acp{cav}, a variant of \ac{mapf} that is considered in this article, considers a dynamic environment by frequently updating control inputs and takes the kinodynamic constraints of vehicles into account during planning.
\Ac{mapf} problems can be solved in a centralized fashion, where all information required to solve every agent's planning problem, such as their objective, state, and constraints, is present in a central computation entity, and this entity computes the motion plan for each agent.
Alternatively, they can be solved in a distributed fashion without a centralized computation entity, where each agent computes only its own motion plan and knows only about its own objective, but communicates with other agents.
\Ac{mamp} in environments of large extent, such as road networks, which typically lack the infrastructure of centralized computation entities, thus requires distributed formulations.
This setting is illustrated in \cref{fig:application} with three \acp{cav} on an intersection.
A more detailed comparison of \ac{mapf} and \ac{mamp} is given in \cref{sec:motion_planning}.

When solving multi-agent planning problems, agents may have coupled objectives or be constrained by each other.
Therefore, the decisions of agents are interdependent.
Solving \iac{mapf} problem optimally by choosing all decision variables simultaneously is NP-hard \citep{yu2013structure,yu2016intractability}.
\Ac{pp}, introduced by \citet{erdmann1987multiple}, is a computationally efficient method for \ac{mapf}.
Instead of solving \iac{mapf} problem optimally, \ac{pp} divides the problem into multiple subproblems, and each agent solves only its subproblem.
Since the subproblems are smaller in size, the computation time is reduced.
\Ac{pp} is particularly well suited to distributed planning problems because it both decomposes the \ac{mapf} problem as described above and follows a sequential, non-iterative scheme that requires only sparse communication.

\subsection{Motivation}

\begin{figure}
    \centering
    \begin{subfigure}[t]{\linewidth}
        \centering
        \includegraphics{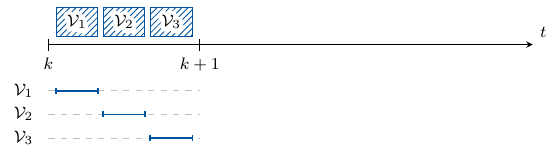}
        \caption{Solving the \ac{pp} problem with a single prioritization \citep[e.g.,][]{kloock2019distributed,wu2020multirobot,vandenberg2005prioritized,yao2018resolving,chalaki2022priorityaware}.}
        \Description{
            Timing diagram over one time step, from time k to time k plus one.
            Above the time axis, a single row of three hatched blocks labeled with agent classes one, two, and three from left to right gives the prioritization, that is, the order of computation.
            Below the time axis, three dashed rows, one per agent class, each contain a single short thick line marking that agent class's computation interval.
            The three intervals lie in three consecutive, non-overlapping slots, so each agent computes in only one third of the available time and idles for the remaining two thirds.
        }
        \label{fig:motivation:single}
    \end{subfigure}
    \par
    \vspace{\baselineskip}
    \begin{subfigure}[t]{\linewidth}
        \centering
        \includegraphics{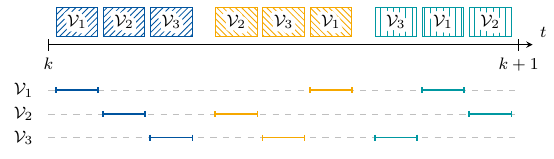}
        \caption{
            Solving the \ac{pp} problem with three prioritizations sequentially \citep[e.g.,][]{bennewitz2002finding,ma2019searching,chan2023greedy}.
            Note how the order of computation changes with the prioritization.
        }
        \Description{
            Timing diagram over the same time step, from time k to time k plus one.
            Above the time axis, nine hatched blocks in a single row form three groups of three, one group per prioritization: first agent classes one, two, three; then two, three, one; then three, one, two.
            Below the time axis, three dashed rows, one per agent class, together contain nine thick lines, each aligned under its block, so that the three prioritizations are computed one after the other.
            Every agent computes three times, but the whole time step is now filled and the total computation time is three times as long as with a single prioritization.
        }
        \label{fig:motivation:sequential}
    \end{subfigure}
    \par
    \vspace{\baselineskip}
    \begin{subfigure}[t]{\linewidth}
        \centering
        \includegraphics{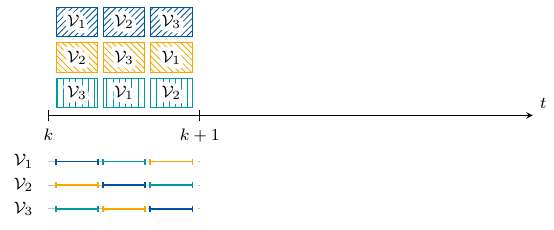}
        \caption{
            Solving the \ac{pp} problem with three prioritizations simultaneously (our approach).
            Note how rows represent the different prioritizations from \cref{fig:motivation:sequential}, and each agent solves only one agent planning problem corresponding to a different prioritization at a time.
        }
        \Description{
            Timing diagram over the same time step, from time k to time k plus one.
            Above the time axis, nine hatched blocks are arranged in a three-by-three grid whose rows are the three prioritizations of the previous subfigure, each row drawn in its own hatching pattern: agent classes one, two, three; two, three, one; and three, one, two.
            Below the time axis, three dashed rows, one per agent class, each contain three thick lines that fill all three time slots, with the hatching pattern of each line indicating which prioritization it belongs to.
            Every slot of every agent is occupied, so all three prioritizations are computed simultaneously within the same time step that a single prioritization needed, and no agent idles.
        }
        \label{fig:motivation:parallel}
    \end{subfigure}
    \caption{
        Exemplary illustration of the computation time in different prioritization approaches in an example with three agents represented in $\classAgents[1]$, $\classAgents[2]$, and $\classAgents[3]$;
        blocks above the time axis indicate the prioritization, i.e., the order of computation of agents;
        thick lines below the time axis indicate the computation time of the respective agents for the respective prioritization.
    }
    \label{fig:motivation}
\end{figure}

While efficient, \ac{pp} often produces suboptimal solutions and is not guaranteed to find a solution at all.
Let $\fcnObjectiveNcs[\fcnPrio]$ denote the summed cost of all agents' optimal motion plans given a prioritization $\fcnPrio$.
A motion plan is optimal given a prioritization if there is no other motion plan with equal cost that results in a motion plan of lower-priority agents with lower cost.
In \ac{mamp}, there is typically a single motion plan with lowest cost due to the asymmetry of the environment; in \ac{mapf}, which often exhibits symmetry and therefore admits multiple plans with the same cost, there exist algorithms to find optimal plans given a prioritization \citep{morag2024prioritised}.
\begin{definition}
    The optimal prioritization $\fcnPrio^{*}$ is the prioritization which yields the solution with highest quality, i.e., with lowest summed cost of all agents' optimal motion plans given the prioritization,
    \begin{equation}
        \fcnPrio^{*} = \argmin_{\fcnPrio} \fcnObjectiveNcs[\fcnPrio] .
    \end{equation}
\end{definition}
If we view prioritization and planning as a two-stage optimization, the solution given a prioritization can be seen as a local optimum.
The solution given the optimal prioritization is denoted as the global optimum in \ac{pp}, which notably can be different from the optimal solution to the multi-agent planning problem.
A core challenge in \ac{pp} is finding a prioritization that results in a solution of high quality.

In general, for $\numAgents$ agents, there exist up to $\numAgents!$ prioritizations.
To approach the global optimum in \ac{pp}, traditional approaches rely on heuristics to improve the solution quality \citep[e.g.,][]{scheffe2022increasing,wu2020multirobot,kloock2019distributed,yao2018resolving,chen2023deep}.
Such heuristic approaches require domain-specific knowledge, making them unsuitable for generalized application.
Further, their performance can be subpar, as heuristics are often tailored towards specific planning problems.

In domain-independent approaches, there is usually no a-priori knowledge about which prioritization yields a close-to-optimal solution.
Thus, the prioritization must be optimized, which involves computing multiple solutions using different prioritizations.
However, computing multiple solutions sequentially might be intractable for real-time applications with constrained computation time.
Consequently, a simultaneous computation of multiple solutions is needed.

\subsection{Contribution}

In \ac{pp}, solution quality depends on the chosen prioritization.
Approaching the global optimum therefore requires evaluating multiple prioritizations, which is often infeasible under real-time constraints in distributed \ac{mamp}.

This article presents a method that exploits idle time induced by sequential \ac{pp} to evaluate multiple prioritizations simultaneously.
Our main contributions are:
\begin{enumerate*}[label=(\roman*)]
    \item a distributed algorithm for simultaneous evaluation of multiple prioritizations, with the same planning complexity per cycle as single-prioritization \ac{pp};
    \item a prioritization update rule that can maintain a well-performing prioritization across successive time steps.
\end{enumerate*}
Besides these main contributions, we also extend the completeness and P-solvability definitions to time-variant prioritizations, and give a graph-theoretic upper bound on the number of distinct prioritizations that is tighter than the worst-case bound of $\numAgents!$ for sparse coupling graphs, with $\numAgents$ being the number of agents.

\Cref{fig:motivation} illustrates our main idea in an example with three interacting, sequentially computing agents.
Blocks display the prioritization, thick lines represent the computation time of agents.
\Cref{fig:motivation:single} shows how approaches based on heuristics solve the \ac{pp} problem for only a single prioritization.
The solution quality strongly depends on the heuristic.
\Cref{fig:motivation:sequential} shows how the \ac{pp} problem can be solved multiple times sequentially regarding multiple prioritizations.
Even though it can improve solution quality, it prolongs the computation time.
In this work, we make use of the fact that in sequential computation the agents idle most of the time: in the example shown in \cref{fig:motivation:single}, agents idle two-thirds of the available time.
\Cref{fig:motivation:parallel} showcases our approach:
By utilizing the idle time, we solve the \ac{pp} problem regarding the three prioritizations from \cref{fig:motivation:sequential} simultaneously.

\paragraph{Scope and assumptions.}
This article focuses on improving solution quality given short computation time in distributed \ac{mamp}.
Accordingly, our evaluation isolates the planning component and does not model communication latency or packet loss.
To isolate the effect of simultaneous prioritization without conflating it with network effects, we assume reliable, low-latency communication.
This assumption is based on ongoing developments in communication technology targeting reliable, low-latency communication, such as 5G ultra-reliable and low-latency communications (URLLC)~\citep[e.g.,][]{li20185g}.
A full end-to-end evaluation that explicitly includes communication time remains an important extension, but is outside the scope of this article.

We also consider only environments with exclusively \acp{cav}, i.e., without uncontrolled vehicles.
This restriction matches controlled environments such as warehouses or fully autonomous road traffic.

\subsection{Related Work}

This section reviews approaches from the literature mapped to a distributed computation architecture in which each agent computes its respective motion plan.
We illustrate the computations of the literature in \cref{fig:motivation}, which displays the computations of three interacting, sequentially computing agents.
The blocks represent a prioritization, i.e., an order of computations, while the thick lines display the computation time.

When searching a prioritization of high solution quality, heuristics are often consulted.
Heuristics aim to improve the prioritization mostly for their domain-dependent objective, and are thus less likely to generalize well.
The following approaches use objective-based heuristics for prioritization to improve the solution quality.
As illustrated in \cref{fig:motivation:single}, these approaches solve the \ac{pp} problem using a single prioritization.
In \citep{wu2020multirobot}, the prioritization is determined by the number of possible paths a robot can take to reach its goal.
The lower this number is, the higher a robot's priority.
Thereby, the probability for each robot to find a solution is meant to be increased, which directly affects the solution quality.
In \citep{vandenberg2005prioritized}, robots with longer estimated travel distances receive higher priority to more evenly distribute the travel time among robots.
\citet{zhang2022learning} proposes a prioritization algorithm based on machine learning, which performs competitively compared to heuristic algorithms.
Our previous work \citep{kloock2019distributed} prioritizes agents with the goal of increasing the traffic flow rate at a road intersection through a heuristic based on the remaining time before an agent enters the intersection.
In \citep{yao2018resolving}, automated vehicles approaching an intersection are prioritized using \ac{MIP}.
Both the prioritization and the vehicle's speed are optimized to minimize the travel time of the vehicles.
The algorithm in \citep{chalaki2022priorityaware} prioritizes vehicles on intersections using job-shop scheduling and thereby reduces the vehicles' average travel time.
Most of the above approaches have in common that the solution for a single prioritization is computed.
Consequently, their computation time is low, as the complexity is
linear in the number of agents, $\bigO(\numAgents)$.
However, the solution quality can be lower than with a different prioritization, or they can fail to find a solution, even though one exists with a different prioritization.

Instead of finding a prioritization heuristically, other works optimize the prioritization.
As illustrated in \cref{fig:motivation:sequential}, these approaches solve the networked planning problem by exploring multiple prioritizations sequentially.
These approaches aim to improve the solution quality by investing computation time.
\citet{bennewitz2002finding} propose a randomized hill-climbing search for finding the optimal prioritization.
Beginning at an initial prioritization, priorities of agents are swapped randomly to increase the solution quality.
Recomputing the solution in each iteration increases the computation effort.
\citet{ma2019searching,chan2023greedy} present an algorithm to lazily explore the space of all prioritizations.
Whenever a conflict between two agents occurs, both possible prioritizations are added to a search tree.
Thus, the computation effort depends on the number of needed iterations until a conflict-free prioritization is found.
The above approaches can compute solutions to multiple prioritizations, and thus find solutions with higher quality than approaches considering only a single prioritization.
However, the solution quality improvement comes at the cost of higher computation time.
For example, the worst-case effort of conflict-based approaches is
$\bigO ( \numAgents\cdot 2^{\numAgents^2} )$,
as for each conflict up to $\numAgents$ plans must be recomputed, and for each of the possible $\numAgents^2$ conflicts two options for prioritization of the involved two agents exist.

Many of the above approaches to \ac{pp} have a centralized computation architecture, i.e., a single, central processing unit computes solutions for all agents in sequence \citep[e.g.,][]{vandenberg2005prioritized,zhang2022learning,ma2019searching,chan2023greedy,yao2018resolving,chalaki2022priorityaware}.
Other centralized approaches explore a larger solution space than in \ac{pp} through a conflict-based search~\citep[e.g.,][]{sharon2015conflictbased,yu2020average,pan2024rhecbs,li2021eecbs,barer2021suboptimal,guo2024expected}.
Although proposed as centralized approaches, they solve local subproblems and could thus be also applied to a distributed computation architecture.
The iterative conflict resolution of the latter group of approaches can result in more sequential computations and communication than in approaches that solve a single prioritization, making them unsuitable when communication resources are scarce.
Even though their computation is efficient in practice, the number of planning iterations is typically unbounded, making them unsuitable for real-time computation.

\Ac{pp} is suited for distributed execution by multiple agents with processing units on each agent~\citep[e.g.,][]{wu2020multirobot,kloock2019distributed,kuwata2007distributed,cap2015prioritized}.
Its non-iterative computation scheme results in scarce usage of communication resources.
These characteristics make it well-suited for the problem setting of \ac{mamp} with distributed computation and constrained computation time of this article.

With multiple processing units, a sequential computation regarding a prioritization yields periods of idle time in each unit.
The problem of reducing idle time due to sequential operation is addressed in production automation.
Johnson's rule \citep{almhanna2023reducing} and sequencing \citep{mak2014sequencing} are methods to decrease idle time by optimizing the assignment and sequence for jobs on machines in assembly lines.
Further, \citet{cam2020formulation} reduce idle time of control systems, i.e., unproductive time of the system, but not idle time during computation.
In this article, we will utilize idle time in sequential computation for simultaneous computation of multiple prioritizations in \ac{pp}, which should increase solution quality without increasing computation time.

\subsection{Structure of This Article}
The rest of this article is organized as follows.
In \cref{sec:preliminaries}, we formalize time-variant coupling and prioritization of agents in a receding horizon planning framework.
In \cref{sec:method}, we formulate requirements for simultaneously computable prioritizations.
We further present our algorithm for simultaneous computation,
which can both retain prioritizations to utilize the predictive nature of receding horizon planning, and explore new prioritizations to increase the likelihood of finding a solution.
In particular, the time required for exploration is spread across consecutive time steps, which keeps computation time at each time step short.
We present our multi-agent motion planner for \acp{cav} driving on roads in \cref{sec:motion_planning}.
In \cref{sec:evaluation}, we present a comparison of our approach to state-of-the-art prioritization approaches in numerical experiments with \acp{cav}, and showcase real-time capabilities of our approach in an experiment with ten \acp{cav} in our \ac{cpmlab}.

\section{Preliminaries}\label{sec:preliminaries}
\begin{figure}
    \centering
    \includegraphics{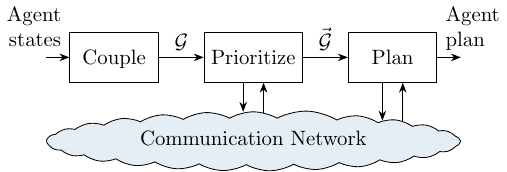}
    \caption{
        The framework for \ac{pp} in which our approach is embedded.
        Each agent builds an undirected coupling graph $\graphUndirected$ which represents potential conflicts,
        negotiates a prioritization with the other agents to build a directed coupling graph $\graphDirected$,
        and plans according to the directed coupling graph.
    }
    \Description{
        Block diagram of the planning framework of a single agent.
        Three boxes are arranged left to right and labeled Couple, Prioritize, and Plan.
        An incoming arrow from the left feeds agent states into Couple.
        An arrow from Couple to Prioritize is labeled with the undirected coupling graph, an arrow from Prioritize to Plan is labeled with the directed coupling graph, and an outgoing arrow to the right emits the agent plan.
        Below the three boxes, a cloud labeled Communication Network is connected by bidirectional arrows to the Prioritize and the Plan box, showing that these two steps exchange information with the other agents.
    }
    \label{fig:preliminaries:framework}
\end{figure}

This section provides the background to prioritized receding horizon planning.
In \ac{pp}, agents first receive all plans from their predecessors, then solve the planning problem, and finally send their plans to their successors.
To determine the predecessors and successors of agents, we couple and prioritize them.
An overview of the \ac{pp} framework is given in \cref{fig:preliminaries:framework}.
In this work, we use exclusively distributed and decentralized algorithms for all components of the framework.
The following subsections introduce our notation and the background on coupling, prioritization, and receding horizon planning.

\subsection{Notation}\label{sec:preliminaries:notation}
In this article, we refer to agents whenever concepts are generally applicable to \ac{pp}.
We assume that states of agents are observable, and therefore we refer to states instead of outputs.
The definitions and methods can be transferred to outputs as well, which we omit for brevity.
We denote
scalars with normal font,
vectors with bold lowercase letters,
matrices with bold uppercase letters, and
sets with calligraphic uppercase letters.
For any set $\mathcal{S}$, the cardinality of the set is denoted by $|\mathcal{S}|$.
A variable $x$ is marked with a superscript $x^{(\anAgent)}$ if it belongs to agent $\anAgent$.
The actual value of a variable $x$ at time $\timestep$ is written as $x({\timestep})$, while values predicted for time $\timestep+\timestepIterator$ at time $\timestep$ are written as $x_{\timestep+\timestepIterator \vert \timestep}$.
A trajectory is denoted by replacing the time argument with $(\cdot)$ as in $x_{\cdot \vert \timestep}$.

This article addresses motion planning for multiple agents.
We refer to the entirety of all agents and the communication links between them as \iac{ncs}, i.e., a control system whose components are connected by a communication network.
We use the prefix \emph{networked} throughout to denote quantities of the \ac{ncs} as a whole, as opposed to those of an individual agent.
We use graphs as a modeling tool of \ac{ncs}.
Every agent is associated with a vertex, so the terms are used synonymously.
\begin{definition}[\Gls{def:graph_undirected}]\label{def:graph_undirected}
    \glsdesc*{def:graph_undirected}
\end{definition}
\begin{definition}[\Gls{def:graph_directed}]\label{def:graph_directed}
    \glsdesc*{def:graph_directed}
\end{definition}

\subsection{Coupling Agents}\label{sec:preliminaries:coupling}

\begin{figure}
    \begin{subfigure}[t]{0.49\linewidth}
        \centering
        \includegraphics{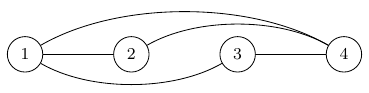}
        \caption{Example undirected coupling graph.}
        \Description{
            Graph with four circular vertices numbered one to four in a horizontal row.
            Undirected edges connect vertex one to vertices two, three, and four, vertex two to vertex four, and vertex three to vertex four.
            Vertices two and three are not connected to each other.
        }
        \label{fig:preliminaries:coupling-graph-undirected}
    \end{subfigure}
    \hfill
    \begin{subfigure}[t]{0.49\linewidth}
        \centering
        \includegraphics{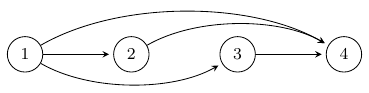}
        \caption{
            Example directed coupling graph.
            An agent can start computation as soon as all its predecessors in the graph have finished their computations and communicated their results.
        }
        \Description{
            The same graph with four circular vertices numbered one to four in a horizontal row, but with every edge now oriented from the lower to the higher vertex number: arrows point from vertex one to vertices two, three, and four, from vertex two to vertex four, and from vertex three to vertex four.
            The result is a directed acyclic graph in which vertex one has no predecessor and vertex four has no successor.
        }
        \label{fig:preliminaries:coupling-graph-directed}
    \end{subfigure}
    \caption{
        Example undirected coupling graph that is transformed to a directed coupling graph by orienting the edges according to the agents' priorities.
        In the example, agent priority is according to vertex number.
    }
\end{figure}

If agents are related via their objectives or constraints, we speak of coupled agents.
A coupling graph represents this relation between agents.
\begin{definition}[\Gls{def:coupling_graph_undirected}]\label{coupling_graph_undirected}
    \glsdesc*{def:coupling_graph_undirected}
\end{definition}
An example coupling graph with four agents is displayed in \cref{fig:preliminaries:coupling-graph-undirected}.
The application in which \ac{pp} is used determines which coupling objectives and constraints must be considered in an agent's planning problem.
For example, in \ac{mapf}, a constraint is introduced to achieve collision avoidance between two agents.
Such a constraint relates the two agents, so they are coupled.

Early approaches to \ac{pp} coupled all agents, resulting in purely sequential computation~\citep[e.g.,][]{erdmann1987multiple}.
Later approaches reduced average computation time by coupling agents ad hoc during planning whenever a conflict arises, thereby interweaving planning and coupling~\citep[e.g.,][]{cap2015prioritized}.
However, the worst-case computation time remains that of purely sequential computation.
Other approaches determine couplings before planning starts using conservative approximations that are less restrictive than coupling all agents.
For example, in robot applications with a fixed time horizon, the robots' movement range over that horizon can be used to determine couplings \citep{baras2003decentralized,kuwata2007distributed,scheffe2024limiting}.
Alternatively, if robots follow fixed paths, path intersections define couplings between robots.
This is often the case for road vehicles following lanes or for warehouse robots \citep{ma2017lifelong,li2023intersection}.
While pre-planning coupling has a higher expected computation time than ad hoc coupling due to its conservativeness, it has the advantage that the number of required sequential computations is known before planning begins.
This knowledge allows limiting per-agent planning time to enable real-time operation.

\subsection{Prioritizing Agents}\label{sec:preliminaries:prioritization}
In \ac{pp}, we prioritize agents, which results in clear responsibilities considering coupling objectives and constraints during planning \citep{kuwata2007distributed,alrifaee2016coordinated}.
A fixed or time-invariant prioritization function $\fcnPrio\colon\setVertices\to\setNaturalNumbers$ prioritizes every agent.
If $\fcnPrio(\anAgent)<\fcnPrio(\anotherAgent)$, then agent $\anAgent$ has a higher priority than agent $\anotherAgent$.

By prioritizing, we can orient the edges of an undirected coupling graph to form a directed coupling graph.
\begin{definition}[\Gls{def:coupling_graph_directed}]\label{coupling_graph_directed}
    \glsdesc*{def:coupling_graph_directed}
\end{definition}
An edge points towards the vertex with lower priority,
\begin{equation}\label{eq:construction-rule}
    \edgeDirected{\anAgent}{\anotherAgent} \in \setEdgesDirected
    \iff
    \fcnPrio (\anAgent) < \fcnPrio(\anotherAgent)
    \land
    \edgeUndirected{\anAgent}{\anotherAgent}\in \setEdges.
\end{equation}
To relate agents that are coupled only indirectly, we use the notion of a path.
\begin{definition}[\Gls{def:path}]\label{def:path}
    \glsdesc*{def:path}
\end{definition}
The prioritization function thus constitutes a strict partial order $\prec$ on the agent set $\setVertices$, given by the transitive closure of the edge relation of $\graphDirected$,
\begin{equation}
    \anAgent \prec \anotherAgent
    \iff
    \text{there is a path from } \anAgent \text{ to } \anotherAgent \text{ in } \graphDirected.
\end{equation}
The order is partial, as opposed to total, since a relation exists only for agents that are coupled directly or indirectly.
For the strict partial order---and thus the orientation of every edge---to be well-defined, a valid prioritization function needs to assign pairwise different priorities to vertices that are connected by an edge:
\begin{equation}
    \label{eq:valid_priority}
    \fcnPrio(\anAgent)\neq \fcnPrio(\anotherAgent),
    \quad
    \forall \anAgent, \anotherAgent \in \setVertices,
    \forall \edgeUndirected{\anAgent}{\anotherAgent}\in\setEdges, \anAgent \neq \anotherAgent.
\end{equation}
Given the construction rule in \cref{eq:construction-rule}, the directed coupling graph $\graphDirected$ resulting from the undirected coupling graph $\graphUndirected$ and a valid prioritization function $\fcnPrio$ regarding \cref{eq:valid_priority} is a \ac{dag}~\citep{scheffe2025graph}.

\begin{definition}[Distinct prioritizations]\label{def:distinct-prioritizations}
    Two prioritizations are considered distinct if they constitute a different strict partial order, or equivalently, if they induce a different directed coupling graph via \cref{eq:construction-rule}.
\end{definition}
We distinguish prioritizations in this way because prioritizations that induce the same directed coupling graph yield identical planning problems.

Since in \ac{pp}, an agent requires its predecessors' plans to consider coupling objectives and constraints, the agents must solve their planning problems in a sequential order.
This order, as well as the communication links, is deducible from the directed coupling graph by traversing the graph along its edges.
An example directed coupling graph with four agents is displayed in \cref{fig:preliminaries:coupling-graph-directed}.

\subsection{Computation Time Definition}\label{sec:preliminaries:computation-time}
In general, the computation time of \ac{mamp} is the duration from measuring the states until control inputs are determined, which consists of the planning time in a locally distributed setting, and additional communication time in a physically distributed setting.
This article is on prioritization of agents, which influences the required number of sequential computations.
Since, in our problem setting, the planning is computationally demanding, we simplify the communication time to be constant and negligible compared to the planning time.
The computation time $\tCompNcs$ of the \ac{ncs} is thus mainly determined by the time to solve the agents' planning problems.

With the following definitions, we formalize the computation time of the \ac{ncs} based on our previous work \citep{scheffe2025graph}.
\begin{definition}[\Gls{def:graph:degree}]\label{def:graph:degree}
    \glsdesc*{def:graph:degree}
\end{definition}
Let the sources of $\graphDirected$ be the set of vertices without incoming edges
\begin{equation}
    \setVertices_s = \set{i\in \setVertices \mid \vertexInDegree{i} = 0}
\end{equation}
and the sinks be the set of vertices without outgoing edges
\begin{equation}
    \setVertices_t = \set{i\in \setVertices \mid \vertexOutDegree{i} = 0}.
\end{equation}
Note that since the coupling graph is a \ac{dag}, there is at least one source and one sink.
Note further that the number of agent classes $\numLevels$ is the diameter of $\graphDirected$ plus one.
The diameter of a graph is the greatest length of any shortest path (\cref{def:path}) between each pair of vertices $\setVertices \times \setVertices$ that is connected by a path.

To determine the computation time of the \ac{ncs}, we add a virtual source $S$ and a virtual sink $T$ to the set of vertices $\setVertices$ of $\graphDirected$, resulting in the set of vertices $\setVertices^{\prime}$.
We furthermore add edges from the source $S$ to all vertices in $\setVertices_s$,
and from all vertices in $\setVertices_t$ to the sink $T$, resulting in the set of edges $\setEdgesDirected^{\prime}$.

\begin{figure}
    \centering
    \includegraphics{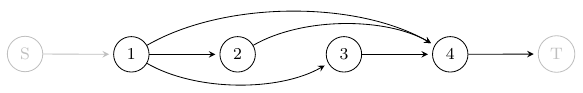}
    \caption{Computation graph for the coupling graph in \cref{fig:preliminaries:coupling-graph-directed}.}
    \Description{
        The directed coupling graph of the previous figure, extended by two auxiliary vertices drawn in light gray: a source vertex S to the left of the row and a sink vertex T to the right.
        A gray arrow leads from S to vertex one, the only vertex without predecessors, and a gray arrow leads from vertex four, the only vertex without successors, to T.
        The black arrows between vertices one to four are unchanged, so that every path from S to T corresponds to a chain of sequential computations.
    }
    \label{fig:preliminaries:computation-graph}
\end{figure}

To obtain the computation time $\tCompNcs$, we weight edges by a weighting function $f_w \colon \setVertices^{\prime} \times \setVertices^{\prime} \to \setRealNumbers$
\begin{equation}\label{eq:weight_func}
    f_w \bigl( \edgeDirected{i}{j} \bigr) =
    \begin{cases}
        \tComp^{(i)}, & \text{ if } \anAgent \in \setVertices, \\
        0,            & \text{ otherwise},
    \end{cases}
\end{equation}
with the computation time $\tComp^{(i)}$ of an edge's respective starting vertex.
The result is the computation graph $\graphDirected^{\prime} = (\setVertices^{\prime}, \setEdgesDirected^{\prime}, f_w)$.
\Cref{fig:preliminaries:computation-graph} illustrates a computation graph.
Let $P_{\max}\subseteq\setEdgesDirected^{\prime}$ denote the longest path between $S$ and $T$ in $\graphDirected^{\prime}$.
The weight of this path corresponds to the computation time of the \ac{ncs}
\begin{equation}\label{eq:computation_time}
    \tCompNcs = \sum_{\edgeDirected{i}{j}\in P_{\max}} f_w \left(\edgeDirected{i}{j}\right).
\end{equation}

\subsection{Receding Horizon Planning with Priorities}\label{sec:preliminaries:rhc}

Traditional \ac{mapf} is a multi-agent planning problem in which agents move in an environment represented by a graph $\graphMap=(\setVerticesMap,\setEdgesMap)$, consisting of a set of vertices $\setVerticesMap$, which are locations, and a set of edges $\setEdgesMap$, which are paths between locations.
The objective for each agent $\anAgent$ is to move quickly from a start vertex $s_{\anAgent} \in \setVerticesMap$ to a target vertex $t_{\anAgent} \in \setVerticesMap$.
The constraints for each agent are to avoid collisions with other agents.
A collision occurs when two agents $\anAgent$ and $\anotherAgent$ at a time $\timestep$ are located at the same vertex $\aVertex$, represented by a tuple $\tuple{\anAgent, \anotherAgent, \aVertex, \timestep}$,
or when they traverse the same edge $\edgeUndirected{\aVertex}{\anotherVertex}\in\setEdgesMap$, represented by a tuple $\tuple{\anAgent, \anotherAgent, \aVertex, \anotherVertex, \timestep}$.

A strategy to reduce computation time in \ac{mapf} is to introduce a fixed time horizon.
Instead of solving the \ac{mapf} problem for the optimal sequence of actions from start to end, only a certain number of time steps $\horizonPrediction$ is considered.
From the sequence of optimal actions for the fixed time horizon of duration $\horizonPrediction \cdot \tSample$, agents apply only the first.
The problem is solved repeatedly after a time step duration $\tSample$ over the course of time.
This practice is known as receding horizon planning or online replanning \citep{silver2005cooperative,li2021lifelong,scheffe2023receding,shahar2021safe}.

In the context of \ac{pp}, priorities can be reassigned in each time step of receding horizon planning, which results in a time-variant prioritization.
We redefine the time-invariant prioritization function to a time-variant prioritization function.
It is a function
$\fcnPrio \colon \setVertices \times \setNaturalNumbers \to \setNaturalNumbers$
which prioritizes each agent $\anAgent$ at each time step $\timestep$.
We indicate a time-invariant prioritization by replacing the time argument with a dash, as in $\fcnPrio(\anAgent,-)$.
In our approach, the priority ordering is consistent during planning for the planning horizon, but can change whenever agents plan.

\begin{figure}
    \centering
    \includegraphics{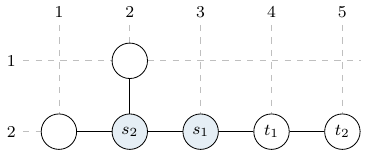}
    \caption{
        A \ac{mapf} instance which is not TP-solvable.
    }
    \Description{
        Grid graph of two rows and five columns with dashed grid lines and numbered rows and columns.
        Six vertices are drawn as circles and connected by edges into a corridor: the bottom row holds vertices at columns one to five, and one additional vertex sits above the corridor in row one, column two, forming a dead end branching off the vertex in row two, column two.
        Two agents are placed in the corridor: agent one starts at row two, column three and must reach its target at row two, column four; agent two starts at row two, column two and must reach its target at row two, column five.
        The start vertices are shaded.
        Both agents must pass through the same narrow corridor in the same direction, and the only way to let agent two pass is for agent one to step aside into the dead-end vertex above, which is a detour away from its own target.
    }
    \label{fig:mapf-example-not-tp-solvable}
\end{figure}

In accordance with the usage in the literature \citep[e.g.,][]{ma2019searching,wu2020multirobot}, we define completeness as follows.
\begin{definition}[Completeness]\label{def:completeness}
    A \acl*{pp} method is complete if, given an underlying planning method, any prioritization results in a solution for every \ac{mapf} instance.
\end{definition}
We extend the considerations in \citep{ma2019searching} from fixed prioritizations to time-variant prioritizations.
\begin{theorem}\label{thm:pp_incomplete}
    In general, \acl*{pp} with time-variant prioritization is incomplete.
\end{theorem}
\begin{proof}
    A counterexample is given in \cref{fig:mapf-example-not-tp-solvable}.
    This example requires the highest-priority agent to take a suboptimal action with regard to its own objective in order
    to create a solution to the \ac{mapf} problem.
    The highest-priority agent, however, always takes an optimal action in prioritized planning.
\end{proof}

\begin{definition}[P-solvable, according to \citep{ma2019searching}]
    A \ac{mapf} instance is P-solvable iff there exists a solution for some fixed prioritization $\fcnPrio(\anAgent,-)$.
\end{definition}
\begin{definition}[TP-solvable]
    A \ac{mapf} instance is TP-solvable iff there exists a solution for some time-variant prioritization $\fcnPrio(\anAgent,\timestep)$.
\end{definition}
\begin{theorem}
    The class of TP-solvable \ac{mapf} instances includes the class of P-solvable \ac{mapf} instances.
\end{theorem}
\begin{proof}
    A time-variant prioritization can mimic a fixed prioritization by using the same ordering at every time step,
    \begin{equation}
        \fcnPrio(\anAgent,\timestep) := \fcnPrio(\anAgent,-) .
    \end{equation}
    Therefore, every \ac{mapf} instance that is solvable with a fixed prioritization is solvable with a time-variant prioritization.
\end{proof}

\begin{figure}
    \centering
    \includegraphics{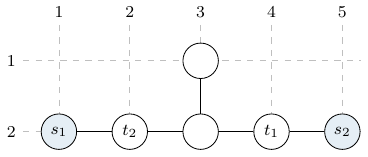}
    \caption{
        A \ac{mapf} instance which is not P-solvable \citep{ma2019searching}, but is TP-solvable with a priority flip in the third time step.
        Figure redrawn from \citep{ma2019searching}.
    }
    \Description{
        Grid graph of two rows and five columns with dashed grid lines and numbered rows and columns.
        Six vertices are drawn as circles: the bottom row holds vertices at columns one to five, connected into a corridor, and one additional vertex sits above the corridor in row one, column three, forming a dead end branching off the middle vertex of the corridor.
        The two agents start at opposite ends of the corridor and must swap sides: agent one starts at row two, column one, shaded, and must reach its target at row two, column four; agent two starts at row two, column five, shaded, and must reach its target at row two, column two.
        Since the corridor is only one vertex wide, one agent must wait in the dead-end vertex above the middle while the other passes, which requires the priority between the two agents to be flipped in the third time step.
    }
    \label{fig:mapf-example-time-variant}
\end{figure}

\begin{theorem}
    Prioritized planning with a time-variant prioritization is incomplete for the class of TP-solvable \ac{mapf} instances.
\end{theorem}
\begin{proof}
    For the \ac{mapf} instance displayed in \cref{fig:mapf-example-time-variant}, a time-variant prioritization that does not result in a solution is any fixed prioritization.
\end{proof}

\section{Exploring Multiple Prioritizations Simultaneously}
\label{sec:method}

\newcommand{\indexPermutation}{q}

This section presents the core idea of this article, which is to improve the solution quality compared to a given prioritization by exploring multiple other prioritizations simultaneously, as illustrated in \cref{fig:motivation:parallel}.
In the \ac{pp} framework (\cref{fig:preliminaries:framework}), our approach replaces both the elements \enquote{Prioritize} and \enquote{Plan}, as we will identify multiple prioritizations for which agents can plan simultaneously.
We formalize parallelizable agent computations by the introduction of agent classes in \cref{sec:method:parallelizable}.
We present in \cref{sec:method:improve-quality} conditions to identify multiple prioritizations which can be computed simultaneously by these agent classes, which allows us to improve solution quality without increasing computation time.
Building on these conditions, we present the algorithm for selecting a set of prioritizations and simultaneously computing with these prioritizations in \cref{sec:method:simultaneous}.
In \cref{sec:method:initial-prioritization}, we show theoretically that our approach can both retain prioritizations to utilize the predictive nature of receding horizon planning and explore new prioritizations to increase the likelihood of finding a solution.

\subsection{Computation Sequences of Agent Classes}
\label{sec:method:parallelizable}
In this section, we summarize how computations of agents can be executed in parallel as introduced in~\citep{alrifaee2016coordinated}.
With the notion of agent classes to describe agents that can compute in parallel, we describe the computation order of agent classes with computation sequences and link computation sequences to prioritization functions.

\begin{figure}
    \centering
    \includegraphics{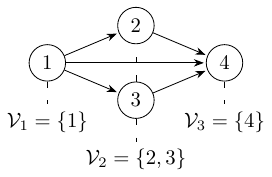}
    \caption{Example coupling graph for \cref{alg:agent-classes}, and resulting agent classes in sequence.}
    \Description{
        Directed coupling graph with four circular vertices numbered one to four.
        Vertex one is on the left, vertices two and three are stacked above and below each other in the middle, and vertex four is on the right.
        Arrows point from vertex one to vertices two, three, and four, and from vertices two and three to vertex four; vertices two and three are not connected.
        Below the graph, dashed lines group the vertices into three agent classes in sequence: the first class contains agent one, the second class contains agents two and three, and the third class contains agent four.
    }
    \label{fig:find_agent_classes}
\end{figure}

The computation order of agents in \ac{pp} can be structured with agent classes $\setClasses$.
Agents belonging to the same agent class can solve their planning problem in parallel.
An undirected coupling graph $\graphUndirected = (\setVertices,\setEdges)$ and a prioritization function $\fcnPrio$ form a directed coupling graph $\graphDirected$ with \cref{eq:construction-rule}.
By topologically sorting the directed coupling graph, we map agents to classes $\setClasses$ according to our \cref{alg:agent-classes}.
This algorithm is executed by every agent simultaneously.
\begin{algorithm}
    \caption{Find agent classes, adapted from \citet{alrifaee2016coordinated}}
    \label{alg:agent-classes}
    \begin{algorithmic}[1]
    \Require Directed coupling graph $\graphDirected = (\setAgents, \setEdgesDirected)$
    \Ensure  Agent classes in sequence $\sequenceClasses$

    \State
        $\numLevels \gets 0$
        \Comment{number of classes}
    \State $\setAgents_\text{todo} \gets \setAgents$ \Comment{remaining agents}
    \While{$\setAgents_\text{todo} \neq \emptyset$}
        \State $\numLevels \gets \numLevels+1$
        \State $\classAgents[\numLevels] \gets \emptyset$
        \For{$i \in \setAgents_\text{todo}$ \label{alg:pv:loop_rem}}
            \If{
                $\vertexInDegree{i} = 0$
                \label{alg:pv:no_inc_edge}
            }
                \State
                    $\classAgents[\numLevels] \gets \classAgents[\numLevels] \cup i$
                    \label{alg:pv:place_v}
                    \Comment{sources can compute}
            \EndIf
        \EndFor
        \If{$\classAgents[\numLevels] = \emptyset$}
            \State \Return FAILURE \Comment{Graph is no DAG}
        \EndIf
        \State $s_{\numLevels} \gets \classAgents[\numLevels]$ \label{alg:pv:store}
        \State $\setAgents_\text{todo} \gets \setAgents_\text{todo} \setminus \classAgents[\numLevels]$ \label{alg:pv:rm_done}
        \For{$i \in \classAgents[\numLevels]$}\label{alg:pv:loop-edges-1}
            \For{$j \in \setAgents_\text{todo}$}\label{alg:pv:loop-edges-2}
                \State
                    $\setEdgesDirected \gets \setEdgesDirected \setminus \edgeDirected{i}{j}$
                    \label{alg:pv:rm-edge}
            \EndFor
        \EndFor
    \EndWhile
    \State \Return $\sequenceClasses$
\end{algorithmic}

\end{algorithm}
The algorithm inspects all vertices to find sources, i.e., vertices with no incoming edges (\cref{alg:pv:no_inc_edge}).
All sources can start their computation, so we add them to a class (\cref{alg:pv:place_v}).
The resulting class is next in solving its planning problem, so it is added to the computation sequence (\cref{alg:pv:store}).
The processed vertices are removed from the set of vertices to process (\cref{alg:pv:rm_done}).
Finally, all edges connected to the processed vertices are removed from the coupling graph (\cref{alg:pv:loop-edges-1,alg:pv:loop-edges-2,alg:pv:rm-edge}).
The loop repeats until all vertices are assigned to classes.
The output of \cref{alg:agent-classes} is a sequence of agent classes $\sequenceClasses$ which orders the agent classes $\setClasses = \set{\classAgents[1], \ldots, \classAgents[\numLevels]}$.
This computation sequence represents the order in which the agent classes can solve their planning problems.
\begin{definition}[Computation sequence]
    \label{def:sequence}
    The set of agent classes $\setClasses$ is a countable set with $\abs{\setClasses} = \numLevels$.
    The computation sequence defined as
    \begin{equation}
        \sequenceClasses = \left(\sequenceElement{1},\sequenceElement{2},\dots,\sequenceElement{\numLevels}\right)
    \end{equation}
    with the domain $\set{1, \ldots, \numLevels}$ and the codomain $\setClasses$ represents a permutation of the agent classes in $\setClasses$.
\end{definition}
\Cref{fig:find_agent_classes} shows an example of \cref{alg:agent-classes} for the directed coupling graph with four agents given in \cref{fig:preliminaries:coupling-graph-directed}.
The resulting computation sequence is
$\bigl( \set{1}, \set{2,3}, \set{4} \bigr)$.

\Cref{alg:agent-classes} returns a computation sequence with the minimum number of sequential computations for the given directed coupling graph.
If we can invert this process, i.e., generate a valid prioritization corresponding to a computation sequence, we can abstract the exploration of different prioritizations to the exploration of different computation sequences.
W.l.o.g., agents in different classes have different priorities,
\begin{equation}
    \fcnPrio(\anAgent) \neq \fcnPrio(\anotherAgent),
    \quad
    \forall \classAgents[1],\classAgents[2]\in\setClasses, \classAgents[1]\neq\classAgents[2],
    \forall \anAgent\in\classAgents[1], \anotherAgent\in\classAgents[2].
\end{equation}
According to \cref{eq:valid_priority}, agents in the same class may have equal priorities as they are not connected by an edge.
However, since we allow time-variant coupling graphs, we enforce unique priorities among agents (cf. \cref{sec:method:initial-prioritization}).
A valid prioritization corresponding to the computation sequence is given by
\begin{equation}\label{eq:prio-from-sequence}
    \fcnPrio(\anAgent) = Z \cdot \numAgents + i, \quad \anAgent \in \sequenceElement{Z} \in \sequenceClasses
    ,
\end{equation}
with an agent $\anAgent$, and the sequence index $Z$.
For the example in \cref{fig:find_agent_classes}, this function would yield
\begin{equation}
    \fcnPrio(1) = 5  , \quad
    \fcnPrio(2) = 10 , \quad
    \fcnPrio(3) = 11 , \quad
    \fcnPrio(4) = 16 .
\end{equation}
With \cref{eq:prio-from-sequence}, we can express the search for a prioritization of high solution quality as the search of a computation sequence permutation of high solution quality.

\subsection{Parallelizable Prioritizations}
\label{sec:method:improve-quality}
The problem of finding the optimal permutation of agent classes can formally be stated as follows.
\begin{problem}
    \label{problem:optimal-sequence}
    Given a set of agent classes $\setClasses$, find a computation sequence ${\sequenceClasses}^*$ resulting in a prioritization function $\fcnPrio_{\indexPermutation}$ with \cref{eq:prio-from-sequence} that minimizes the sum of the costs of all agents
    \begin{equation}
        \label{eq:problem-optimal-sequence}
        \fcnPrio_{\indexPermutation} = \argmin_{\fcnPrio} \fcnObjectiveNcs[\fcnPrio].
    \end{equation}
\end{problem}
Note that finding the optimal sequence does not necessarily coincide with finding the optimal prioritization, as the permutations of sequences may not express the permutations of prioritizations.

\cref{problem:optimal-sequence} is known as the Drilling Problem, which is a problem in combinatorial optimization \citep[p.1]{korte2018combinatorial}.
Given a set of holes that need to be drilled, the Drilling Problem consists of finding a permutation of the order in which they need to be drilled to minimize the total distance the drill needs to travel.
As the set of classes contains $\lvert\setClasses\rvert = \numLevels$ elements, for \cref{problem:optimal-sequence}, this would result in $\lvert\setSequences\rvert=\numLevels!$ different permutations.
A possible solution to that problem is enumerating all $\numLevels!$ sequences \citep{korte2018combinatorial,azarm1997conflictfree}:
Given a set $\setClasses$ of $\numLevels$ classes, solve \ac{pp} problems regarding each of the $\numLevels!$ sequences and choose the one with minimal cost.
However, the computation time of this approach increases rapidly with the number of classes.

Solving \cref{problem:optimal-sequence} is computationally intractable with real-time constraints.
According to the goal of this article to increase solution quality without increasing computation time, we aim to solve the following problem.
\begin{problem}
    \label{problem:set-permutations}
    Given a set of agent classes $\setClasses$, find $\numLevels = \lvert\setClasses\rvert$ different computation sequences such that all sequences can be computed simultaneously.
\end{problem}
Our goal in \cref{problem:set-permutations} is to find $\numLevels$ sequences out of $\numLevels !$ that can be computed simultaneously in a distributed setting.
A class $\classAgents[l] \in \setClasses$ can solve exactly one planning problem regarding a specific computation sequence in each time slot.
Note that in a centralized setting any computation sequence can be computed simultaneously, given enough computational resources.
However, this is only possible since all information required to solve every agent's planning problem, such as their objective, state, and constraints, is present in the central entity.

To formalize the task, we introduce a computation schedule matrix $\matSchedule\in\setClasses^{\numLevels\times\numLevels}$.
Each cell contains an agent class, and each row contains a computation sequence.
Each column of $\matSchedule$ represents a time slot in the networked computation.
The computation schedule matrix thus defines for each agent the prioritization for which it should plan: an agent looks up its agent class and finds the corresponding prioritization, i.e., the computation sequence, in the matrix.
The matrix $\matSchedule$ represents a valid computation schedule if its entries fulfill
\begin{align}
     & \matScheduleElement{ij} \neq \matScheduleElement{ik},
     & i,j,k \in\set{1,\dots,\numLevels}, \, j \neq k , \text{ and}
    \label{eq:row-distinct}                                         \\
     & \matScheduleElement{ij} \neq \matScheduleElement{kj},
     & i,j,k \in\set{1,\dots,\numLevels}, \, i \neq k .
    \label{eq:column-distinct}
\end{align}
We ensure valid computation sequences with \cref{eq:row-distinct},
and we ensure that each class solves exactly one planning problem in each time slot with \cref{eq:column-distinct}.
The matrix resulting from these conditions is also known as a Latin square.
Adapted from \citet{mckay2005number}, a Latin square $\bm{L} \in \set{1,\dots,N}^{N \times N}$ is a matrix in which each row and column contains only distinct entries.
That is, each row and column contains every element in $\set{1,\dots,N}$ exactly once.
For $\matSchedule$, the property of distinct entries in each row and column is given by \cref{eq:row-distinct} and \cref{eq:column-distinct}, respectively.
The problem is related to a Sudoku puzzle: the solution to a Sudoku puzzle is a Latin square with $N=9$ with the additional rule that each of the nine non-overlapping $3\times3$ partial matrices contains every element in $\set{1,\dots,N}$ exactly once.

\begin{figure}
    \centering
    \begin{subfigure}[t]{0.24\linewidth}
        \centering
        \includegraphics{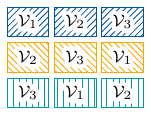}
        \caption{Example computation schedule matrix.}
        \Description{
            Three-by-three matrix of hatched blocks, each row drawn in its own hatching pattern and color.
            The rows are cyclic shifts of the three agent classes: the first row contains agent classes one, two, three; the second row two, three, one; and the third row three, one, two.
            Every agent class therefore appears exactly once in each row and exactly once in each column.
        }
        \label{fig:method:computation-graph-schedule}
    \end{subfigure}
    \hfill
    \begin{subfigure}[t]{0.74\linewidth}
        \centering
        \includegraphics{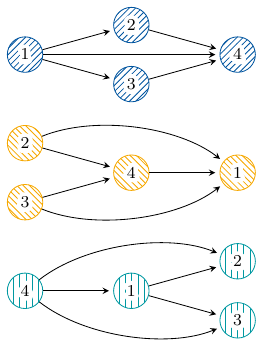}
        \caption{
            Corresponding directed coupling graphs; vertices are arranged in rows and columns of the computation schedule matrix.
            Edges in black between agents indicate coupling.
            An agent can start computation as soon as all its predecessors in the graph have finished their computations and communicated their results.
        }
        \Description{
            Three directed coupling graphs stacked vertically, one per row of the computation schedule matrix, drawn in the hatching pattern and color of that row.
            Each graph contains the same four agents, numbered one to four, and the same five black coupling edges, but the edges are oriented differently in each graph because the agents are ordered by a different prioritization.
            The vertices of each graph are placed in the columns that correspond to the time slot in which that agent computes for that prioritization, so that in the top graph agent one computes first, in the middle graph agents two and three compute first, and in the bottom graph agent four computes first.
        }
        \label{fig:method:coupling-graph}
    \end{subfigure}
    \caption{
        Example computation schedule matrix and computation graph for our explorative prioritization.
        Four agents in three agent classes simultaneously compute three different prioritizations.
        Agent classes are from the example in \cref{fig:find_agent_classes}.
    }
    \label{fig:method:computation-graph}
\end{figure}

An example computation schedule matrix and the corresponding coupling graphs are displayed in \cref{fig:method:computation-graph}.
Starting from the prioritization and resulting agent classes of the example in \cref{fig:find_agent_classes}, a valid computation schedule matrix is presented in \cref{fig:method:computation-graph-schedule}.
Note how each agent class appears exactly once per column (at the same time, an agent class can only do one computation) and once per row (each computation sequence permutation requires one computation per agent class).
The three coupling graphs corresponding to the computation schedule matrix are given in \cref{fig:method:coupling-graph}.
With three different coupling graphs, three different solutions are computed simultaneously.

\subsection{Simultaneous Computation with Multiple Prioritizations}\label{sec:method:simultaneous}

This section describes our algorithms for simultaneous computation with multiple prioritizations.
We first present our decentralized algorithm through which every agent obtains the same computation schedule matrix which fulfills \cref{eq:column-distinct,eq:row-distinct}.
We then show how agents solve \ac{pp} problems with multiple prioritizations.
Finally, we show how agents select a solution for the \ac{pp} problems.

Our goal for the computation schedule matrix $\matSchedule$ is to explore $\numLevels$ prioritizations in a time step to increase the probability of finding the optimal prioritization according to \cref{problem:optimal-sequence}.
\Cref{alg:computation-schedule} is our decentralized algorithm to find a computation schedule matrix $\matSchedule$, which is executed by each agent simultaneously.
We restate that a row $\matScheduleVector{\indexPermutation\cdot}\in\setClasses^{1 \times \numLevels}$ corresponds to a computation sequence $\indexPermutation$,
and a column $\matScheduleVector{\cdot m}\in\setClasses^{\numLevels \times 1}$ corresponds to a time slot $m$ during the networked computation.
First, we initialize the computation schedule matrix $\matSchedule$ with the size $\numLevels \times \numLevels$ and fill the first row $\matScheduleVector{1\cdot}$ with the classes of $\setClasses$ in sequence (\cref{alg:cs:init-schedule,alg:cs:first-seq}).
Note that this initial row can change from one time step to another as it depends on the initial prioritization.
We enforce a common random seed among agents for consistent results of the algorithm (\cref{alg:cs:seed}).
The algorithm determines the remaining rows in a loop (\cref{alg:cs:loop-seq}).
For each row, the algorithm loops over all $\numLevels$ columns that need to be populated (\cref{alg:cs:n-slots,alg:cs:loop-slots}).
We use the number of available classes in each column that can form a valid computation schedule matrix as a heuristic for the order in which the columns are populated (\cref{alg:cs:find-options}).
This number of available classes is stored in the \code{options} array.
It is an array of sets, each containing all elements of $\setClasses$ that respect \cref{eq:row-distinct} regarding the current row $\indexPermutation$ and that respect \cref{eq:column-distinct} regarding the column corresponding to the array index.
The next column to populate is the one with the least number of available classes (\cref{alg:cs:find-slot}).
One of these classes is selected at random (\cref{alg:cs:random-option}).
If all columns are populated with valid entries, we progress to the next row (\cref{alg:cs:loop-seq-progress}).
If there is a column for which no classes are available to select from, the current row is incompatible with the rest of the computation schedule matrix (\cref{alg:cs:invalid}), so we reset the current row (\cref{alg:cs:reset-seq}) and start over.
\begin{algorithm}
    \caption{Decentralized construction of computation schedule matrix $\matSchedule$}
    \label{alg:computation-schedule}
    \begin{algorithmic}[1]
    \Require Set of agent classes $\setClasses= \set{\classAgents[1], \classAgents[2], \ldots, \classAgents[\numLevels]}$
    \Ensure  Computation schedule matrix $\matSchedule$

    \State $\matSchedule \gets \text{empty matrix of size } \numLevels \times \numLevels$
        \label{alg:cs:init-schedule}
    \State $\matScheduleVector{1\cdot} \gets [ \classAgents[1], \classAgents[2], \ldots, \classAgents[\numLevels] ]$
        \label{alg:cs:first-seq}

    \State $q \gets 1$

    \State initialize random seed with current time step \Comment{for consistency across agents}
        \label{alg:cs:seed}

    \While{$q < \numLevels$ }
        \Comment{find remaining columns}
        \label{alg:cs:loop-seq}

        \State \code{is\_valid} $\gets$ \textbf{true}

        \State $N_{\text{cols-todo}} \gets \numLevels$
            \label{alg:cs:n-slots}

        \While{$N_{\text{cols-todo}} > 0$}
            \label{alg:cs:loop-slots}

            \For{$ m = 1, \ldots, \numLevels $}
                \State \code{options[$m$]} $\gets \setClasses \setminus \left( \matScheduleVector{q \cdot} \cup \matScheduleVector{\cdot m} \right)$
                    \label{alg:cs:find-options}
            \EndFor

            \State $m \gets \argmin_{i} \vert$\code{options[$i$]}$\vert$
                \label{alg:cs:find-slot}

            \If{$\vert$\code{options[$i$]}$\vert = 0$}
                \label{alg:cs:invalid}

                \State \code{is\_valid} $\gets$ \textbf{false}
                \State \textbf{break}
            \EndIf

            \State $\matScheduleElement{qm} \gets$ \Call{Random}{\code{options[$m$]}}
                \label{alg:cs:random-option}

            \State $N_{\text{cols-todo}} \gets N_{\text{cols-todo}} - 1$
        \EndWhile

        \If{\code{is\_valid}}
            \State $q \gets q + 1$
                \label{alg:cs:loop-seq-progress}
        \Else
            \State $\matScheduleVector{q\cdot} \gets \text{empty vector of size } 1 \times \numLevels$
                \label{alg:cs:reset-seq}
        \EndIf

    \EndWhile
    \State \Return $\matSchedule$
\end{algorithmic}

\end{algorithm}

\begin{figure}
    \centering
    \includegraphics{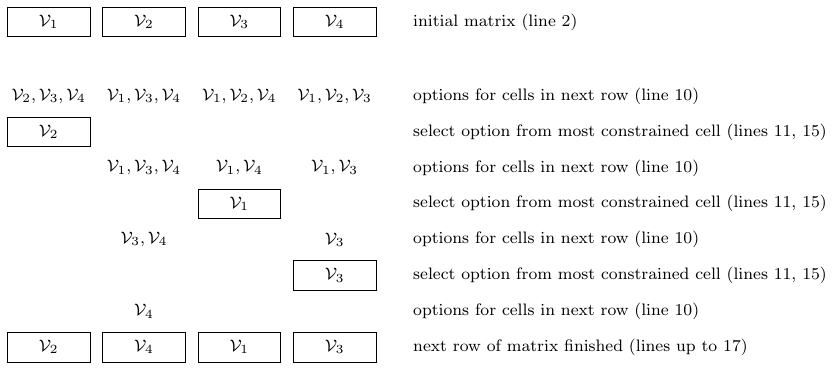}
    \caption{Example for \cref{alg:computation-schedule} which builds a computation schedule matrix.}
    \Description{
        Step-by-step trace of the computation schedule algorithm for four agent classes, drawn as a sequence of table rows with four columns.
        The top row is the initial matrix row containing agent classes one, two, three, and four in order.
        Each following pair of rows shows first the remaining options for every column, then the option that is selected from the most constrained column, which is framed.
        The options shrink from three candidates per column to two, then to one, as classes two, one, three, and finally four are fixed.
        The bottom row shows the finished next matrix row, containing agent classes two, four, one, and three.
        Annotations on the right of each row name the corresponding lines of the algorithm.
    }
    \label{fig:computation-schedule}
\end{figure}

We illustrate \cref{alg:computation-schedule} with an example in \cref{fig:computation-schedule}.
The figure contains four agent classes, and depicts the construction of the second row of the computation schedule matrix.
Each other row in the figure corresponds either to determining the remaining options of computation classes for each cell (\cref{alg:cs:find-options}), or to the random selection of an option (\cref{alg:cs:random-option}) from the most constrained cell (\cref{alg:cs:find-slot}).

Our \cref{alg:computation-schedule} builds a computation schedule row by row, which is valid according to \cref{eq:row-distinct,eq:column-distinct}.
A partly built, valid computation schedule can always be completed.
Therefore, our \cref{alg:computation-schedule} will eventually converge to a valid computation schedule.
Due to the randomness of the approach, there is no bound on the number of required iterations until convergence, which could be addressed with a more sophisticated algorithm.
In practice, the computation time to build a computation schedule matrix is negligible when compared to the computation time for motion planning.

Computation with multiple prioritizations is similar in concept to computation with a single prioritization.
The main difference is that there are $\numLevels$ different directed coupling graphs $\graphDirected_{\fcnPrio_{\indexPermutation}}=(\setVertices,\setEdgesDirected_{\fcnPrio_{\indexPermutation}})$, and with it sets of
predecessors $\setPredecessors{\anAgent}_{\fcnPrio_{\indexPermutation}}$,
successors $\setSuccessors{\anAgent}_{\fcnPrio_{\indexPermutation}}$, and
neighbors $\setNeighbors{\anAgent}_{\fcnPrio_{\indexPermutation}}$, defined as
\begin{gather}
    \setPredecessors{\anAgent}_{\fcnPrio_{\indexPermutation}} = \set{\anotherAgent\in\setVertices \mid \edgeDirected{\anotherAgent}{\anAgent}\in\setEdgesDirected_{\fcnPrio_{\indexPermutation}}},
    \\
    \setSuccessors{\anAgent}_{\fcnPrio_{\indexPermutation}} = \set{\anotherAgent\in\setVertices \mid \edgeDirected{\anAgent}{\anotherAgent}\in\setEdgesDirected_{\fcnPrio_{\indexPermutation}}},
    \\
    \setNeighbors{\anAgent}_{\fcnPrio_{\indexPermutation}} = \setPredecessors{\anAgent}_{\fcnPrio_{\indexPermutation}} \cup \setSuccessors{\anAgent}_{\fcnPrio_{\indexPermutation}}
    .
\end{gather}
Each agent performs $\numLevels$ computations in $\numLevels$ time slots.
An agent in class $\classAgents[l]$ finds the sequence $\matScheduleVector{\indexPermutation\cdot}$ corresponding to a time slot $m$ by looking for its agent class in the respective column of the computation schedule matrix.
For this computation sequence it must hold
\begin{equation}
    \matScheduleElement{\indexPermutation{}m} = \classAgents[l]
    \,.
\end{equation}
From this sequence, the agent constructs the directed coupling graph $\graphDirected_{\fcnPrio_{\indexPermutation}}$ with \cref{eq:prio-from-sequence,eq:construction-rule}.
The agent solves its planning problem after receiving the required information from its predecessors $\setPredecessors{\anAgent}_{\fcnPrio_{\indexPermutation}}$.
It then stores the solution cost corresponding to the sequence $\indexPermutation$ as an element in a solution cost array.

After all computation sequences are solved, each agent broadcasts its solution cost array.
With this information, each agent can determine the solution quality of every computation sequence.
The solution quality $\fcnObjectiveNcs(\timestep)$ of a time step $\timestep$ for each computation sequence $\indexPermutation$ is measured by the networked cost: the sum of the costs of all agent planning problems regarding the corresponding computation sequence
\begin{equation}
    \label{eq:networked_cost_timestep}
    \fcnObjectiveNcs[\fcnPrio_{\indexPermutation}](\timestep)
    = \sum_{\anAgent=1}^{\numAgents} \fcnObjective[\anAgent]_{\fcnPrio_{\indexPermutation}}(\timestep),
\end{equation}
with the cost $\fcnObjective[\anAgent]_{\fcnPrio_{\indexPermutation}}\colon\setNaturalNumbers\to\setRealNumbers$ of an agent $\anAgent$,
the sequence ${\indexPermutation}$ corresponding to a prioritization function $\fcnPrio_{\indexPermutation}$ with \cref{eq:prio-from-sequence},
and a time step $\timestep$.
After receiving the cost arrays of all other agents, the value for the solution quality for each computation sequence is identical in each agent.
Each agent selects the computation sequence with the highest solution quality, i.e., the minimal networked cost
\begin{equation}
    \label{eq:select-prioritization}
    \fcnPrio_{\text{explore}} = \argmin_{\fcnPrio_{\indexPermutation}} \fcnObjectiveNcs[\fcnPrio_{\indexPermutation}](\timestep)
    .
\end{equation}
In case two prioritizations yield the same cost, the one with the lower index is chosen by the algorithm.

\subsection{Initial Prioritization}\label{sec:method:initial-prioritization}

Our goal of improving the solution quality compared to a given prioritization implies the assumption that an initial valid prioritization according to \cref{eq:valid_priority}, i.e., a prioritization constituting a strict partial order, is given.
Finding an initial valid prioritization is simple, e.g., a prioritization that assigns priorities according to unique vertex numbers is valid.
Other prioritization algorithms exist \citep{kuwata2007distributed,scheffe2025graph,scheffe2022increasing}.
The selection of the initial prioritization, and with \cref{alg:agent-classes} the initial computation sequence as well, determines which prioritizations can possibly be explored.
We identified two objectives for selecting the initial prioritization.

The time-variance of priorities in receding horizon planning has both advantages and disadvantages.
An advantage of time-variance is that situations which dictate varying priorities over time to find a solution can be solved, as illustrated by \cref{fig:mapf-example-time-variant}.
A disadvantage is that the constraints in an agent's planning problem change with the prioritization.
In a receding horizon planning algorithm, an agent's control inputs are optimized while adhering to constraints in the future, i.e., the time horizon.
A change in the constraints can counteract the predictive nature of a receding horizon planning algorithm, which can affect the solution quality \citep{scheffe2025graph}.
For example, if an agent must come to a full stop during the time horizon to avoid collision with a higher-priority agent, it might decelerate.
However, if the priorities flip and the other agent avoids the collision, deceleration reduces solution quality unnecessarily.
Therefore, one objective is to enable the algorithm to retain priorities, and thus avoid it being forced to change priorities.
The other objective is to explore many prioritizations to find the best available solution.
A poor selection of an initial set of agent classes can limit the exploration of prioritizations to a fraction of the number of distinct prioritizations.

We achieve the first objective by using the computation sequence with the lowest cost according to \cref{eq:networked_cost_timestep} from a previous time step for the initial prioritization.
From this computation sequence, we generate priorities with \cref{eq:prio-from-sequence}.
To guarantee the validity of these priorities according to \cref{eq:valid_priority}, they must be unique to account for the time-variability of the coupling graph.
In the next time step, we orient the undirected coupling graph by prioritizing agents with the retained priorities, thus creating a directed coupling graph.
From this directed coupling graph, we construct the initial computation sequence with our \cref{alg:agent-classes}.
With this process, it is always possible to maintain the priorities of the previous time step even with a time-variant coupling graph.

We took a step towards the second objective by random construction of the computation schedule matrix in \cref{alg:computation-schedule}.
However, given a computation sequence, only a subset of computation schedule matrices can be constructed.
The question remains: which computation sequences can be explored given an initial sequence?
We explore different computation sequences if we build unique computation schedule matrices.
Two computation schedule matrices are distinct if the sets of their rows are different, i.e., if one matrix contains a computation sequence that the other one does not contain.
In other words, they are the same if one can be mapped to the other by rearranging its rows.
It is easy to see that for $\numLevels = 1$ and $\numLevels = 2$, we obtain the optimal computation sequence.
There exists only a single computation schedule matrix, so all sequences can be explored.
For $\numLevels = 3$, there exist two unique computation schedule matrices which do not share a single row.
With the initial prioritization selected by retaining priorities, we can transition from one computation schedule matrix to another between time steps as long as they share an identical row, i.e., an identical computation sequence.
Consequently, retaining the result for an initial prioritization for $\numLevels = 3$ restricts us to examining only three out of six possible computation sequences.
More formally, we can construct a computation schedule graph
$\graphUndirected_{c} = \left(\setVertices_{c}, \setEdges_{c} \right)$
with the set of all unique computation schedule matrices as vertices $\setVertices_{c}$, which are connected by an edge if they share an identical row
\begin{equation}
    \label{eq:computation-schedule-graph-edges}
    \edgeUndirected{i}{j} \in \setEdges_{c}
    \iff
    \exists n,m \colon \matScheduleVector{n \cdot}^{(i)} = \matScheduleVector{m \cdot}^{(j)}
\end{equation}
with
the row $\matScheduleVector{n \cdot}^{(i)}$ belonging to the computation schedule matrix in vertex $i$, and
the row $\matScheduleVector{m \cdot}^{(j)}$ belonging to the computation schedule matrix in vertex $j$.

\begin{theorem}
    \label{th:computation-schedule-connectivity}
    Let the initial prioritization be selected by retaining priorities, and let the computation schedule matrix be created with \cref{alg:computation-schedule}.
    For $\numLevels \geq 4$, we can explore all possible computation sequences of length $\numLevels$ given enough time.
\end{theorem}
\begin{proof}
    We can prove this theorem by showing that the computation schedule graph $\graphUndirected_{c}$ contains a spanning tree.
    Let an ascending sequence denote a sequence in which the indices of the agent classes are consecutive for consecutive elements, except for the index $\numLevels$, which is followed by $1$.
    We prove our theorem by showing that all computation schedule matrices are connected to a computation schedule matrix which contains the ascending sequence $(\classAgents[1], \ldots, \classAgents[\numLevels])$.
    All matrices containing the computation sequence $(\classAgents[1], \ldots, \classAgents[\numLevels])$ are connected according to \cref{eq:computation-schedule-graph-edges}.
    All $\numLevels$ different ascending sequences fit into one computation schedule matrix by shifting the starting index by one for each row.
    It follows that if any ascending sequence is part of a computation schedule matrix, the computation sequence $(\classAgents[1], \ldots, \classAgents[\numLevels])$ is part of a connected matrix.
    If the initial sequence does not allow any ascending sequence to be part of the computation schedule matrix, exactly one element of this initial sequence is responsible for making exactly one of the ascending sequences invalid.
    Put differently: as long as there are two agent classes with consecutive indices in the initial sequence---both of which belong to a single ascending sequence and thus invalidate the same ascending sequence---one ascending sequence must be valid in the corresponding computation schedule matrix.
    In the case that the initial sequence does not allow any ascending sequence to be part of the computation schedule matrix, we can place any ascending sequence and switch the element that would make the sequence invalid with its following element.
    This results in a sequence where $\numLevels - 2$ elements are consecutive in their agent class index.
    If $\numLevels \geq 4$, more than two elements of this computation sequence are consecutive in their agent class index.
    As pointed out above, this matrix is connected to a matrix which contains a valid ascending sequence, which in turn is connected to a matrix containing the sequence $(\classAgents[1], \ldots, \classAgents[\numLevels])$.
    Therefore, a spanning tree of $\graphUndirected_{c}$ is rooted in the corresponding vertex.
\end{proof}

\subsection{Discussion of Our Approach for Simultaneous Computation with Multiple Prioritizations}
\label{sec:method:discussion}
The number of prioritizations that our approach considers equals the number of agent classes, which is upper bounded by the number of agents.
A commonly referenced upper bound for the number of distinct prioritizations is $\numAgents!$, which is the number of total orders on the set of agents, i.e., the number of permutations of the set of agents.
Consequently, this bound guarantees only a ratio of explored prioritizations to distinct prioritizations that decreases rapidly with the number of agents.
However, since only a partial order is required, the number of distinct prioritizations is related to the connectivity of the coupling graph, which is captured by the vertex degrees $\vertexDegree{i}$ (\cref{def:graph:degree}).
\begin{theorem}\label{th:number-of-prioritizations}
    The number of distinct prioritizations $N_{\text{prio}}$ of an undirected coupling graph $\graphUndirected = (\setVertices, \setEdges)$ is upper bounded by
    \begin{equation}
        N_{\text{prio}}
        \leq
        \min \set{ \prod_{i\in\setVertices} \left( \vertexDegree{i} + 1 \right), \; \numAgents ! }
        .
    \end{equation}
\end{theorem}
\begin{proof}
    A valid prioritization function orients every edge of $\graphUndirected$ with \cref{eq:construction-rule}, and the resulting directed coupling graph is acyclic (\cref{sec:preliminaries:prioritization}).
    Conversely, every acyclic orientation of $\graphUndirected$ is induced by the prioritization that assigns priorities according to a topological order of that orientation.
    With \cref{def:distinct-prioritizations}, $N_{\text{prio}}$ thus equals the number of acyclic orientations of $\graphUndirected$.
    The first bound is the upper bound on the number of acyclic orientations given in \citep[Theorem 10]{manber1984effect}.
    The second bound follows since each such topological order is one of the $\numAgents!$ total orders on $\setVertices$.
\end{proof}
\begin{remark}
    According to \cref{th:number-of-prioritizations}, the bound on the number of distinct prioritizations decreases as the coupling graph becomes sparser, i.e., as the vertex degrees decrease.
    On sparse coupling graphs, our approach therefore explores a larger fraction of the distinct prioritizations than the bound $\numAgents!$ suggests.
\end{remark}
Our approach only explores a fraction of all distinct prioritizations.
However, if the networked planning problem changes only slightly from one time step to another, or if the approach is applied repeatedly, our approach can improve the solution quality incrementally.
A significant benefit of our approach is that it is general and does not require any domain-specific knowledge.

\section{Motion Planning for CAVs}\label{sec:motion_planning}

\Ac{mamp} is a variant of \ac{mapf}.
The evaluation of the approach presented in this article takes place in the context of \ac{mamp} for \acp{cav}.
Extending the description of the problem setting in \cref{sec:introduction}, we state the following differences between traditional \ac{mapf} and our formulation of \ac{mamp} for \acp{cav}.
\begin{enumerate}
    \item \emph{System dynamics}:
          In \ac{mapf}, the dynamics of robots are usually abstracted in the graph search problem \citep[e.g.,][]{banfi2020planning,li2021lifelong}.
          In our \ac{mamp}, we explicitly consider the system dynamics in the planning problem by encoding motion in \iac{mpa}.
    \item \emph{Objective}:
          In \ac{mapf}, the objective of agents is to plan a path to reach a target vertex in the graph representing the environment without collisions.
          In our \ac{mamp}, the objective of agents is to plan motions to stay close to a reference path without collisions.
          The reference path is the result of a routing layer and does not consider obstacles.
    \item \emph{Planning time horizon}:
          In \ac{mapf}, the planning time horizon is variable and extends until the agent has reached its target vertex.
          An agent plans once, and the plan reaches from the agent's start vertex to its target vertex.
          In our \ac{mamp}, the planning time horizon is fixed.
          This allows for short and predictable computation times.
          An agent plans at every time step, thereby shifting its planning horizon, an approach known as \ac{rhc}.
    \item \emph{Environment representation}:
          In \ac{mapf}, the environment with its obstacles is encoded in a graph, where vertices represent locations at which agents can stop, and edges represent discretized motions of agents.
          In our \ac{mamp}, the environment with its obstacles is represented by a list of polygons.
          Each agent is required to avoid these polygons when planning motions.
\end{enumerate}

There are works that go beyond traditional \ac{mapf} and extend the above-listed categorization; e.g., system dynamics are considered in \citep{alonso-mora2018cooperative,luo2016distributed,le2018cooperative}.

We formulate motion planning as \iac{ocp} in \cref{sec:mamp-mp}.
We present our model for considering kinodynamic constraints in \cref{sec:mamp-mpa}, and we show how we couple agents in \cref{sec:mamp-couple}.
\Cref{sec:mamp-mpc} presents our algorithm to solve the planning problem of each agent, which is based on \ac{mpc}.

\subsection{Motion Planning as an Optimal Control Problem}\label{sec:mamp-mp}

This section presents the \acp{ode} describing the vehicle dynamics and our cost function before both are incorporated into \iac{ocp} for motion planning.
The variables and equations in this section belong to a single agent $\anAgent$.

As introduced in \cref{sec:preliminaries:notation}, we refer to the entirety of all agents as \iac{ncs}.
We control multiple agents in \iac{ncs} with a combination of prioritized planning and \ac{dmpc} \citep{richards2007robust}.
This combination will be referred to as \ac{pdmpc} \citep{scheffe2022increasing,scheffe2024prioritized,scheffe2024limiting,kloock2019networked}.
Finding the agents' control inputs is modeled as \iac{ocp}.
We denote by agent \acp{ocp} the \acp{ocp} solved by the agents,
and by networked \ac{ocp} the \ac{ocp} for the entire prioritized \ac{ncs}.

\begin{figure}
    \centering
    \includegraphics{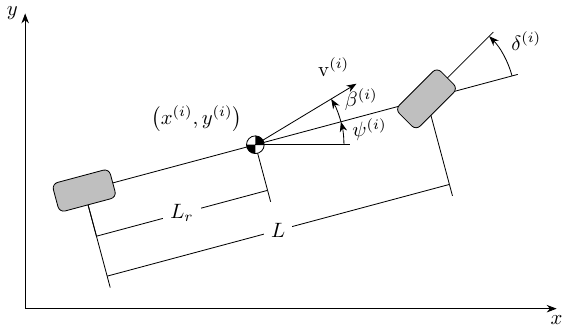}
    \caption{Kinematic single-track model of a vehicle. Figure redrawn from \citep{scheffe2023receding}.}
    \Description{
        Sketch of the kinematic single-track model in a Cartesian coordinate system with a horizontal x axis and a vertical y axis.
        The vehicle is drawn as a line between a rear and a front wheel, both shown as rounded gray rectangles.
        The rear wheel is aligned with the vehicle body, while the front wheel is turned by the steering angle delta, marked as an angle between the body axis and the wheel.
        The center of gravity, marked by the usual black-and-white quartered circle, lies on the body line and is labeled with its position coordinates x and y.
        The yaw angle psi is the angle between the x axis and the vehicle body, and the side slip angle beta is the angle between the vehicle body and the velocity vector, which is drawn as an arrow from the center of gravity and labeled with the speed.
        Dimension lines below the vehicle give the distance from the rear wheel to the center of gravity and the total wheelbase.
    }
    \label{fig:kst_model}
\end{figure}
\Cref{fig:kst_model} shows an overview of variables for the nonlinear kinematic single-track model \citep[p. 21]{rajamani2006vehicle}.
Assuming low velocities, we model no slip on the front and rear wheels, and no forces acting on the vehicle.
The resulting equations are
\begin{equation} \label{eq:kst_model}
    \left\lbrace\hspace{1ex}
    \begin{aligned}
        \dot{x}\ofAgent{\anAgent}(t)         & = \sysSpeed\ofAgent{\anAgent}(t) \cdot \cos(\psi\ofAgent{\anAgent}(t) + \beta\ofAgent{\anAgent}(t)),                         \\
        \dot{y}\ofAgent{\anAgent}(t)         & = \sysSpeed\ofAgent{\anAgent}(t) \cdot \sin(\psi\ofAgent{\anAgent}(t) + \beta\ofAgent{\anAgent}(t)),                         \\
        \dot{\psi}\ofAgent{\anAgent}(t)      & = \sysSpeed\ofAgent{\anAgent}(t) \cdot \frac{1}{L} \cdot \tan(\delta\ofAgent{\anAgent}(t)) \cos(\beta\ofAgent{\anAgent}(t)), \\
        \dot{\sysSpeed}\ofAgent{\anAgent}(t) & = \sysInSpeed\ofAgent{\anAgent}(t),                                                                                          \\
        \dot{\delta}\ofAgent{\anAgent}(t)    & = \sysInSteering\ofAgent{\anAgent}(t),
    \end{aligned}
    \right.
\end{equation}
with
\begin{equation} \label{eq:beta}
    \beta\ofAgent{\anAgent}(t) = \tan^{-1}\left(\frac{L_r}{L} \tan(\delta\ofAgent{\anAgent}(t)) \right),
\end{equation}
where
$x\ofAgent{\anAgent}\in\setRealNumbers$ and $y\ofAgent{\anAgent}\in\setRealNumbers$ describe the position of the \ac{cg},
$\psi\ofAgent{\anAgent}\in[0,2\pi)$ is the orientation,
$\beta\ofAgent{\anAgent}\in[-\pi,\pi)$ is the side slip angle,
$\delta\ofAgent{\anAgent}\in[-\pi,\pi)$ and $\sysInSteering\in\setRealNumbers$ are the steering angle and its derivative, respectively,
$\sysSpeed\ofAgent{\anAgent}\in\setRealNumbers$ and $\sysInSpeed\in\setRealNumbers$ are the speed and acceleration of the \ac{cg}, respectively,
$ L $ is the wheelbase length and $L_r$ is the length from the rear axle to the \ac{cg}.

The system dynamics defined in \cref{eq:kst_model} are compactly written as
\begin{equation}\label{eq:dynamics-general-continuous}
    \dot{\sysState}\ofAgent{\anAgent}(t) := \frac{d}{dt} \sysState\ofAgent{\anAgent} (t) = f \big( \sysState\ofAgent{\anAgent}(t), \sysControlInputs\ofAgent{\anAgent}(t) \big)
\end{equation}
with the state vector
\begin{equation}
    {\sysState\ofAgent{\anAgent}} = \begin{pmatrix} x\ofAgent{\anAgent} & y\ofAgent{\anAgent} & \psi\ofAgent{\anAgent} & \sysSpeed\ofAgent{\anAgent} & \delta\ofAgent{\anAgent} \end{pmatrix}\transposed \in\setRealNumbers^5,
\end{equation}
the control input
\begin{equation} \label{eq:input_vector_kst}
    {\sysControlInputs\ofAgent{\anAgent}} = \begin{pmatrix} \sysInSpeed\ofAgent{\anAgent} & \sysInSteering\ofAgent{\anAgent} \end{pmatrix}\transposed \in\setRealNumbers^2
\end{equation}
and the vector field $f$ defined by \cref{eq:kst_model}.
Transferring \cref{eq:dynamics-general-continuous} to a discrete-time nonlinear system representation yields
\begin{equation}\label{eq:dynamics-general-discrete}
    \sysState\ofAgent{\anAgent}_{k+1} = f_d \bigl( \sysState\ofAgent{\anAgent}_{k}, \sysControlInputs\ofAgent{\anAgent}_{k} \bigr)
\end{equation}
with a time step $k\in\setNaturalNumbers$ and the vector field $f_d \colon \setRealNumbers^5 \times \setRealNumbers^2 \to \setRealNumbers^5$.

We define the cost function for agent $\anAgent$ to minimize given a prioritization $\fcnPrio$ in our motion planning problem as
\begin{equation}\label{eq:cost_mp}
    \fcnObjective_{\fcnPrio}\ofAgent{\anAgent}(k) = \sum_{\timestepIterator=1}^{\horizonPrediction}
    \left(
    \sysState\ofAgent{\anAgent}\forTimeAtTime{\timestep+\timestepIterator}{\timestep} -
    \sysRef\ofAgent{\anAgent}\forTimeAtTime{\timestep+\timestepIterator}{\timestep}
    \right)\transposed
    \bm{Q}
    \left(
    \sysState\ofAgent{\anAgent}\forTimeAtTime{\timestep+\timestepIterator}{\timestep} -
    \sysRef\ofAgent{\anAgent}\forTimeAtTime{\timestep+\timestepIterator}{\timestep}
    \right)
\end{equation}
with the prediction horizon $\horizonPrediction\in\setNaturalNumbers$,
a reference trajectory $\sysRef\ofAgent{\anAgent}\forTimeAtTime{\cdot}{\timestep} \in \setRealNumbers^{5}$,
and the positive semi-definite, block diagonal matrix
\begin{equation}
    \bm{Q} =
    \begin{pmatrix}
        \bm{I}_2           & \bm{0}_{2\times 3} \\
        \bm{0}_{3\times 2} & \bm{0}_3           \\
    \end{pmatrix}
    \in \setRealNumbers^{5\times 5} .
\end{equation}

The \ac{ocp} of agent $\anAgent$ follows in \cref{eq:4-ocp}.
Since multiple agents are involved in the equation, we will include the agent superscript.
We assume a full measurement or estimate of the state $\sysState\ofAgent{\anAgent}(\timestep)$ is available at the current time $\timestep$.

\begin{mini!}
{
    \sysControlInputs\ofAgent{\anAgent}\forTimeAtTime{\cdot}{\timestep}
}
{
    \fcnObjective\ofAgent{\anAgent}_{\fcnPrio}(k)
    \label{eq:pdmpc:ocp_obj}
}
{
    \label{eq:4-ocp}
}
{
}
\addConstraint{
    \sysState\ofAgent{\anAgent}\forTimeAtTime{\timestep+\timestepIterator+1}{\timestep}
}
{
    = \sysModelDiscrete \big(
    \sysState\ofAgent{\anAgent}\forTimeAtTime{\timestep+\timestepIterator}{\timestep},
    \sysControlInputs\ofAgent{\anAgent}\forTimeAtTime{\timestep+\timestepIterator}{\timestep}
    \big)
    ,\quad
    \protect\label{eq:pdmpc:ocp_c_system}
}
{
    \timestepIterator=0,\ldots,\horizonPrediction-1
}
\addConstraint{
    \sysState\ofAgent{\anAgent}\forTimeAtTime{\timestep+\timestepIterator}{\timestep}
}
{
    \in \setFeasibleStates
    ,\quad
    \protect\label{eq:pdmpc:ocp_c_state}
}
{
    \timestepIterator=1,\ldots,\horizonPrediction-1
}
\addConstraint{
    \sysState\ofAgent{\anAgent}\forTimeAtTime{\timestep+\horizonPrediction}{\timestep}
}
{
    \in \setFeasibleStates_f
    \protect\label{eq:pdmpc:ocp_c_final_state}
}
\addConstraint{
    \sysState\ofAgent{\anAgent}\forTimeAtTime{\timestep}{\timestep}
}
{
    = \sysState\ofAgent{\anAgent}(\timestep)
    \protect\label{eq:pdmpc:ocp_c_init_state}
}
\addConstraint{
    \sysControlInputs\ofAgent{\anAgent}\forTimeAtTime{\timestep+\timestepIterator}{\timestep}
}
{
    \in \setFeasibleInputs
    ,\quad
    \protect\label{eq:pdmpc:ocp_c_input}
}
{
    \timestepIterator=0,\ldots,\horizonControl-1
}
\addConstraint{
    \fcnConstraintCoupling\ofAgent{\anAgent \seqCoupling \anotherAgent} \left(
    \sysState\ofAgent{\anAgent}\forTimeAtTime{\timestep+\timestepIterator}{\timestep},
    \sysState\ofAgent{\anotherAgent}\forTimeAtTime{\timestep+\timestepIterator}{\timestep}
    \right)
}
{
    \leq 0
    ,\quad
    \protect\label{eq:pdmpc:ocp_c_coupling_sequential}
}
{
    \timestepIterator = 1,\ldots, \horizonPrediction, \quad
    \anotherAgent \in \setPredecessors{\anAgent}_{\fcnPrio}(\timestep)
    .
}
\end{mini!}
Here,
$\sysModelDiscrete \colon \setRealNumbers^{\numStates}\times\setRealNumbers^{\numInputs}\to\setRealNumbers^{\numStates}$ is a vector field representing the discrete-time nonlinear system model with
$\numStates\in\setNaturalNumbers$ as the number of states and $\numInputs\in\setNaturalNumbers$ as the number of inputs,
$\sysControlInputs\ofAgent{\anAgent}\forTimeAtTime{\cdot}{\timestep} = (\sysControlInputs\ofAgent{\anAgent}\forTimeAtTime{\timestep}{\timestep},\sysControlInputs\ofAgent{\anAgent}\forTimeAtTime{\timestep+1}{\timestep}, \dots, \sysControlInputs\ofAgent{\anAgent}\forTimeAtTime{\timestep+\horizonControl-1}{\timestep})$ is the input trajectory,
$\setFeasibleStates$ is the set of feasible states,
$\setFeasibleStatesTerminal$ is the set of feasible terminal states, and
$\setFeasibleInputs$ is the set of feasible inputs.
The coupling constraint $\fcnConstraintCoupling\ofAgent{\anAgent \seqCoupling \anotherAgent} \colon \setRealNumbers^{\numStates} \times \setRealNumbers^{\numStates} \to \setRealNumbers$ in \cref{eq:pdmpc:ocp_c_coupling_sequential} enforces networked constraints among agents.
The set of predecessors $\setPredecessors{\anAgent}_{\fcnPrio}(\timestep)$ contains the neighbors of $\anAgent$ with higher priority given the prioritization $\fcnPrio$.
An agent $\anAgent$ solves its \ac{ocp} after having received the plans of all predecessors $\setPredecessors{\anAgent}_{\fcnPrio}(\timestep)$.
Afterwards, it communicates its own plan to agents in $\setSuccessors{\anAgent}_{\fcnPrio}(\timestep)$.
Since the planning problem requires the predecessors' plans in \cref{eq:pdmpc:ocp_c_coupling_sequential}, the agents must solve their \acp{ocp} in a sequential order.
Each agent solves the \ac{ocp} \cref{eq:4-ocp} repeatedly after a time step duration $\tSample$ and with updated values for the states and constraints, which establishes the \ac{rhc}.
Only the first input $\sysControlInputs[\anAgent]\forTimeAtTime{\timestep}{\timestep}$ of the input trajectory is applied.

\subsection{System Modeling with a Motion Primitive Automaton}\label{sec:mamp-mpa}

This section presents how we model the state-continuous system \cref{eq:dynamics-general-discrete} as \iac{mpa} \citep{frazzoli2005maneuverbased,scheffe2023receding}.
The \ac{mpa} incorporates the constraints on system dynamics \cref{eq:pdmpc:ocp_c_system}, on control inputs \cref{eq:pdmpc:ocp_c_input}, and on both the steering angle and the speed \labelcref{eq:pdmpc:ocp_c_state,eq:pdmpc:ocp_c_final_state}.

\newcommand{\setAutomatonStates}{\mathcal{Q}}
\newcommand{\setAutomatonTransitions}{\mathcal{S}}
\newcommand{\automatonState}{Q}
\newcommand{\automatonTransition}{S}
From the system dynamics \cref{eq:kst_model}, we derive a \ac{fsa} which we call \ac{mpa} and define as follows.
\begin{definition}[\Acl{mpa}, from \citep{scheffe2023receding}]
    An \ac{mpa} is a 5-tuple $(\setAutomatonStates,\setAutomatonTransitions,\gamma,\automatonState_0,\setAutomatonStates_f)$ composed of:
    \begin{itemize}
        \item $\setAutomatonStates$ is a finite set of automaton states $\automatonState$;
        \item $\setAutomatonTransitions$ is a finite set of transitions $\automatonTransition$, also called \acp{mp};
        \item $\gamma:\setAutomatonStates\times \setAutomatonTransitions \times \setNaturalNumbers \to \setAutomatonStates$ is the update function defining the transition from one automaton state to another, dependent on the time step in the horizon;
        \item $\automatonState_0\in \setAutomatonStates$ is the initial automaton state;
        \item $\setAutomatonStates_f\subseteq \setAutomatonStates$ is the set of final automaton states.
    \end{itemize}
\end{definition}
An automaton state is characterized by a specific speed $\sysSpeed$ and steering angle $\delta$.
An \ac{mp} is characterized by an input trajectory and a corresponding state trajectory which starts and ends with the speed and steering angle of an automaton state.
It has a fixed duration which we choose to be equal to the time step duration $\tSample$.
\Acp{mp} can be concatenated to form longer state trajectories by rotation and translation.
Our \ac{mpa} discretizes both the state space with the update function $\gamma$ and the time space with a fixed duration $\tSample$ for all \acp{mp}.
This \ac{mpa} replaces the system representation \cref{eq:dynamics-general-discrete}.
Note that the dynamics model on which our \ac{mpa} is based is interchangeable.
Its complexity is irrelevant computation-wise for motion planning since \acp{mp} are computed offline.
The \ac{mpa} design forces vehicles to come to a complete stop at the end of the prediction horizon, which guarantees that solutions to \cref{eq:4-ocp} are recursively feasible; for details see our previous work \citep{scheffe2022increasing,scheffe2023receding}.

\subsection{Coupling of Vehicles}\label{sec:mamp-couple}

Although coupling all vehicles guarantees that coupling constraints for all vehicles will be considered, it also prolongs computation time and increases the number of prioritizations compared to coupling fewer vehicles.
Therefore, we only couple vehicles that can potentially collide within the horizon $\horizonPrediction$.
We couple two vehicles at a time step $\timestep$ if their reachable sets intersect at at least one time step within the horizon $\horizonPrediction$.
\begin{definition}[Reachable set of a time interval, from \citet{scheffe2024limiting}]\label{def:reachableSet}
    The reachable set $\setReachableB{\anAgent}\forTimeAtTime{[t_1,t_2]}{t_0}$ of the states of vehicle $\anAgent$ between time $t_1$ and time $t_2$ starting from time $t_0$ is
    \begin{equation}\label{eq:reachableSet}
        \setReachableB{\anAgent}\forTimeAtTime{[t_1,t_2]}{t_0} = \biggl\{
        \bigcup_{t' \in [t_1, t_2]} \int_{t_0}^{t'} f(\sysState\ofAgent{\anAgent},\sysControlInputs\ofAgent{\anAgent})dt \nonumber
        \biggm|
        \sysState\ofAgent{\anAgent}(t_0) \in \setFeasibleStates(t_0), \sysControlInputs\ofAgent{\anAgent} \in \setFeasibleInputs
        \biggr\}.
    \end{equation}
\end{definition}
The reachable sets $\setReachableB{\anAgent}\forTimeAtTime{[\timestep+\timestepIterator,\timestep+\timestepIterator+1]}{\timestep}$ over the prediction horizon of a vehicle can be computed offline with the \ac{mpa} presented in \cref{sec:mamp-mpa}.
The set of coupling edges is
\begin{equation}\label{eq:isCoupled}
    \setEdges(\timestep) = \Bigl\{
    \edgeUndirected{\anAgent}{\anotherAgent}
    \Bigm|
    \anAgent, \anotherAgent \in \setAgents,\,
    \anAgent \neq \anotherAgent,\,
    \exists \timestepIterator\in \{0, \ldots, \horizonPrediction-1\}
    \colon \,
    \setReachableB{\anAgent}\forTimeAtTime{[\timestep+\timestepIterator,\timestep+\timestepIterator+1]}{\timestep} \cap
    \setReachableB{\anotherAgent}\forTimeAtTime{[\timestep+\timestepIterator,\timestep+\timestepIterator+1]}{\timestep} \neq
    \emptyset
    \Bigr\}.
\end{equation}
This results in a time-variant coupling graph $\graphUndirected(\timestep)=\left(\setVertices, \setEdges(\timestep)\right)$.
Because the coupling rule in \cref{eq:isCoupled} is deterministic, each agent can compute the coupling graph locally from the current states and reachable-set estimates of all agents.
Hence, all agents obtain the same coupling graph and no central entity is required.
The required state and reachable-set information can be exchanged directly among agents, i.e., in a distributed manner.

\subsection{Planning Using Monte-Carlo Tree Search in a Model Predictive Control Framework}\label{sec:mamp-mpc}

It is computationally hard to find the global optimum to the planning problem \cref{eq:4-ocp} due to its nonlinearity and nonconvexity.
We substitute the system model \cref{eq:pdmpc:ocp_c_system} with \iac{mpa} based on our previous work \citep{scheffe2023receding}.
Our \ac{mpa} is general rather than tailored to a specific environment.
Therefore, our motion planning is a receding horizon tree search rather than a graph search.
Obstacles are included in the state constraints \cref{eq:pdmpc:ocp_c_state} of our \ac{ocp} formulation.
In the \ac{mcts}, we represent obstacles and predecessors' trajectories as polygons.

The cost of edges in the tree is given by the cost function \cref{eq:pdmpc:ocp_obj}.
In this article, we use a sampling-based approach based on \ac{mcts} to find a path in the tree with low cost.
Due to its anytime property, \ac{mcts} is a popular approach to complex control problems \citep{choudhury2022scalable,baby2023monte,kurzer2018decentralized}, as it allows balancing computation effort and solution quality.
In each time step, our \ac{mcts} randomly expands a fixed number of vertices in the search tree.
The algorithm thus inspects only a part of the complete search tree.
After the expansions, the path in the search tree with the lowest cost according to \cref{eq:pdmpc:ocp_obj} is selected.
The algorithm has no guarantee on optimality, but nearly constant computation time.

Because the \ac{mamp} runs online, vehicles require an input at every time step.
However, since \ac{pp} is incomplete (\cref{thm:pp_incomplete}), the \ac{mcts} might fail to find a feasible solution.
In such cases, vehicles reuse their previous, recursively feasible solution, ensuring safe motion plans \citep{scheffe2022increasing}.

\section{Evaluation}\label{sec:evaluation}
We evaluate the presented approach for increasing solution quality by exploring multiple prioritizations in the context of \ac{mamp} for \acp{cav}.
In \cref{sec:evaluation:setup}, we describe the experiment setup, a scenario on a road network with up to 20 vehicles.
In \cref{sec:evaluation:quality,sec:evaluation:time}, we evaluate the quality of the motion plans of the vehicles in the \ac{ncs}, and the computation time of the \ac{ncs}, respectively.
In \cref{sec:evaluation:experiment}, we demonstrate our approach in an experiment in the \ac{cpmlab}, an open-source, remotely accessible testbed for \acp{cav} \citep{kloock2021cyberphysical}.

We compare our prioritization algorithm with five algorithms from the literature, which we selected for their suitability for distributed execution and their similarity in computational complexity.
Each algorithm is represented by a prioritization function $\fcnPrio \colon \setVertices \to \mathbb{N}$.
The first function prioritizes vehicles according to their vertex number \citep{alrifaee2016coordinated} and is denoted by $\fcnPrio_\text{constant}$.
The second function prioritizes vehicles randomly at each time step \citep{bennewitz2002finding} and is denoted by $\fcnPrio_\text{random}$.
The third function prioritizes vehicles with a constraint-based heuristic \citep{scheffe2022increasing}.
A higher number of potential collisions with other vehicles results in a higher priority.
The function is denoted by $\fcnPrio_\text{constraint}$.
The fourth function prioritizes vehicles to decrease the number of agent classes, and thus the computation time, via graph coloring \citep{scheffe2025graph,kuwata2007distributed}.
The function is denoted by $\fcnPrio_\text{color}$.
The fifth function evaluates all distinct prioritizations by solving the networked \ac{ocp} for all acyclic orientations of the coupling graph.
The function then chooses the prioritization which results in the highest solution quality.
It is denoted by $\fcnPrio^{*}$.
In our evaluation, all algorithms above group agents into agent classes for parallel computation as described in \cref{sec:method:parallelizable}.

\subsection{Setup}\label{sec:evaluation:setup}
In our experiments, vehicles move on the road network shown in \cref{fig:evaluation:experiment}.
It consists of an urban intersection with eight incoming lanes, a highway, and highway on- and off-ramps.
With this variety of road segments, we challenge our motion planner with merging, crossing, and overtaking scenarios.

\begin{figure}
    \centering
    \includegraphics{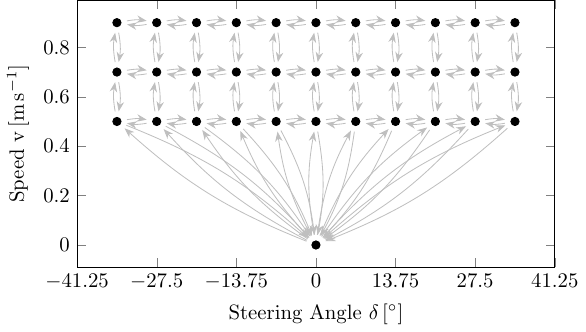}
    \caption{
        The \ac{mpa} used in our experiments.
    }
    \Description{
        Scatter plot of the motion primitive automaton.
        The horizontal axis is the steering angle in degrees, the vertical axis is the speed in meters per second.
        Each marker is a trim state, that is, a combination of steering angle and speed; the trims form a grid-like arrangement with a wide range of steering angles at low speed and a narrowing range at higher speeds.
        Light gray arrows between markers are the transitions of the automaton, connecting each trim only to neighboring trims that are reachable within one time step, so that changes in speed and steering angle are limited.
    }
    \label{fig:evaluation:mpa}
\end{figure}

Our \ac{mcts} uses the \ac{mpa} illustrated in \cref{fig:evaluation:mpa}, which is based on the kinematic single-track model as presented in \citep{scheffe2023receding}.
The \ac{mpa} is discretized with four speed levels.
The number of expansions during low-level planning for a single vehicle is 6000.

We set the duration for our numerical experiments to \qty{7}{\second} with a time step size of \qty{0.2}{\second}.
We evaluate the performance in 50 randomized variants on the road network.
The vehicles start at random positions on the road network and follow a randomly selected path which starts and ends at the same point.
The reference speeds for the vehicles in the experiments are selected at random from the speed levels of our \ac{mpa} shown in \cref{fig:evaluation:mpa}.

During the experiment, the vehicles need to stay within the road boundaries.
Lower-prioritized vehicles must avoid collisions with higher-prioritized ones.
Therefore, coupling constraints with higher-prioritized vehicles reduce the set of feasible system states $\setFeasibleStates$ by the system states the higher-prioritized vehicles occupy during a motion primitive in the \ac{mpa} \citep{scheffe2023receding}.
Two vehicles in \iac{ncs} are coupled whenever their reachable sets overlap \citep{scheffe2024limiting}.
As the reachable set is limited in size, and vehicles move along the road network, couplings between vehicles appear and disappear over the course of an experiment.
Therefore, the coupling graph is time-variant.
We initialize priorities in the very first time step of an experiment with the agents' vertex numbers.

The numerical experiments ran on an off-the-shelf laptop with an Apple M4 Pro 14-Core CPU and \qty{24}{\giga\byte} of RAM using MATLAB 2025b, and the low-level motion planning algorithm implemented in \Cpp.
The physical experiment (\cref{sec:evaluation:experiment}) ran on ten Intel NUCs (NUC7i5BNK), one for each participating vehicle, with an Intel Core i5 7260U CPU with \qty{2.2}{GHz} and \qty{16}{GB} of RAM using MATLAB 2023a, and the low-level motion planning algorithm implemented in MATLAB.
The NUCs are connected via Ethernet, and communication is realized by the ROS~2 Toolbox for MATLAB.
The open-source algorithms implemented in MATLAB and \Cpp are available online%
\footnote{\href{https://github.com/pscheffe/p-dmpc}{github.com/pscheffe/p-dmpc}}.

\subsection{Solution Quality}\label{sec:evaluation:quality}

\begin{figure}
    \centering
    \includegraphics{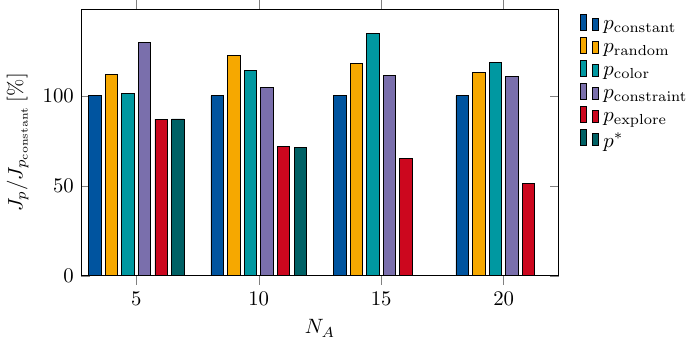}
    \caption{
        Comparison of the networked cost among multiple prioritization algorithms, normalized to the cost of $\fcnPrio_{\text{constant}}$.
        There is no value for $\fcnPrio^{*}$ for 15 and 20 vehicles as the optimization did not terminate.
    }
    \Description{
        Grouped bar chart.
        The horizontal axis is the number of agents with the four groups 5, 10, 15, and 20; the vertical axis is the networked cost in percent of the cost of the constant prioritization.
        Each group holds six bars, one per prioritization algorithm: constant, random, color, constraint, explore, and optimal.
        The constant bar is at 100 percent by definition in every group.
        The random, color, and constraint bars lie above 100 percent, between roughly 101 and 135 percent.
        Our explorative prioritization is the lowest of the heuristic bars in every group and decreases with the number of agents, from about 87 percent at 5 agents to about 72, 65, and 51 percent at 10, 15, and 20 agents.
        The optimal prioritization is close to the explorative one, at about 87 percent for 5 and about 71 percent for 10 agents, and is missing for 15 and 20 agents because the optimization did not terminate.
    }
    \label{fig:evaluation:cost}
\end{figure}

We measure the solution quality with the networked cost $\fcnObjectiveNcs[\fcnPrio]$ of a prioritization $\fcnPrio$.
The networked cost according to the cost of the planning problem of each vehicle is given as
\begin{equation}
    \fcnObjectiveNcs[\fcnPrio]
    = \sum_{\timestep=0}^{\timestep_{\text{experiment}}} \sum_{\anAgent=1}^{\numAgents} \fcnObjective[\anAgent]_{\fcnPrio}(\timestep),
\end{equation}
with $\timestep_{\text{experiment}}$ being the number of time steps in the experiment, and $\fcnObjective[\anAgent]_{\fcnPrio}(\timestep)$ given in \cref{eq:cost_mp}. %
\Cref{fig:evaluation:cost} shows the cost normalized to that of the prioritization $\fcnPrio_\text{constant}$.

Our approach, denoted by $\fcnPrio_{\text{explore}}$, outperforms the approaches from the literature in terms of networked cost across all our experiments.
For five and ten vehicles, the networked cost of our approach is within \qty{1}{\percent} of the cost of the optimal prioritization.
For 15 vehicles, we reduce the networked cost by more than \qty{53}{\percent} compared to the baseline prioritization ($\fcnPrio_{\text{constant}}$).
There is no value for the cost of the optimal prioritization for 15 and 20 vehicles as the algorithm did not terminate in reasonable computation time.

Surprisingly, \cref{fig:evaluation:cost} indicates that constant priorities can yield better solution quality than time-varying priorities even in dynamic environments.
A possible reason is that agents can make full use of the predictive capabilities of \ac{mpc}: the agent \acp{ocp} remain similar between time steps, as opposed to the changing collision avoidance constraints that come with changing priorities.
However, there are instances in which a different prioritization yields better solutions or a different prioritization is required to even enable progress at all.
In contrast to approaches from the literature, our approach can either remain at a prioritization or switch to another one, determined by the solution quality.

At the start of each scenario, vehicles are distributed on the road network, leaving them with enough room to maneuver.
During operation, due to the limited horizon of the planning algorithm, they can end up in standstill situations that cannot be resolved by \ac{pp}, referred to as deadlock.
\Cref{tab:success_rate} displays the success rate of each prioritization, i.e., the ratio of deadlock-free scenario variations to all scenario variations.
For scenarios with five and ten vehicles, all approaches perform similarly and can solve most variants without a deadlock.
For scenarios with 15 and 20 vehicles, however, our proposed approach $\fcnPrio_{\text{explore}}$ exhibits a much higher success rate, cutting deadlock scenarios by up to \qty{88}{\percent}.
\begin{table}
    \centering
    \caption{
        Success rate of evaluated prioritizations, measured as the ratio of the number of deadlock-free scenario variations to all scenario variations.
        Our proposed approach $\fcnPrio_{\text{explore}}$ has the highest success rate compared to other prioritizations.
        There is no value for $\fcnPrio^{*}$ for 15 and 20 vehicles as the optimization did not terminate.
    }
    \label{tab:success_rate}
    \begin{tabular}{@{} r SSSSSS @{}}\toprule
        $\numAgents$ & $\fcnPrio_{\text{constant}}$ & $\fcnPrio_{\text{random}}$ & $\fcnPrio_{\text{color}}$ & $\fcnPrio_{\text{constraint}}$ & $\fcnPrio_{\text{explore}}$ & $\fcnPrio^{*}$
        \\ \midrule
        5            & 1.00                         & 1.00                       & 1.00                      & 0.98                           & 1.00                        & 1.00
        \\
        10           & 0.94                         & 0.94                       & 0.94                      & 0.98                           & 1.00                        & 1.00
        \\
        15           & 0.82                         & 0.88                       & 0.84                      & 0.84                           & 0.98                        & {--}
        \\
        20           & 0.54                         & 0.60                       & 0.62                      & 0.54                           & 0.86                        & {--}
        \\ \bottomrule
    \end{tabular}
\end{table}

If computation time allows, the solution quality of the random prioritization $\fcnPrio_{\text{random}}$ as well as our prioritization $\fcnPrio_{\text{explore}}$ could be improved by computing multiple rounds of \ac{pp} (cf. \cref{sec:conclusion:future-work}).

\subsection{Computation Time}\label{sec:evaluation:time}
This section evaluates the computation time of our approach.
We first extend the definition of the networked computation time to the case of multiple simultaneously computed prioritizations in \cref{sec:evaluation:computation-time-definition}, and then analyze the resulting computation times in \cref{sec:evaluation:time-planning}.

\subsubsection{Computation Time Definition}\label{sec:evaluation:computation-time-definition}

The computation time when planning with a single prioritization was presented in \cref{sec:preliminaries:computation-time}.
Our approach explores one prioritization per agent class.
When exploring multiple prioritizations, the networked computation time is at least as high as the maximum of all considered prioritizations.
It can even be higher, as computations of multiple prioritizations are interdependent: agents need to finish their computations before the next prioritization can be computed.

\begin{figure}
    \centering
    \includegraphics{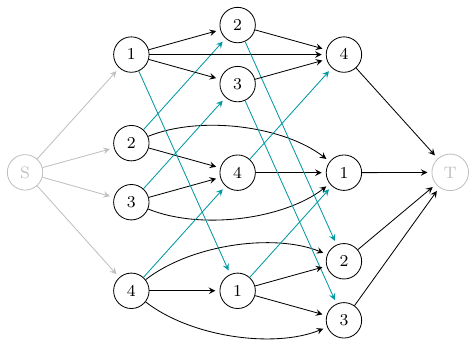}
    \caption{
        Computation graph corresponding to the coupling graphs in \cref{fig:method:coupling-graph}; vertices are arranged in rows and columns of the computation schedule matrix.
        Edges in black between agents indicate coupling.
        Edges in teal are given in addition to coupling, since an agent can only work on one prioritization at a time.
        Edges and nodes in light gray are for determining the networked computation time via a longest path algorithm.
    }
    \Description{
        Computation graph for four agents that simultaneously compute three prioritizations.
        The vertices are the twelve pairs of agent and prioritization, arranged in the rows and columns of the computation schedule matrix, so that the column of a vertex is the time slot in which that agent solves that prioritization.
        Black edges within each of the three blocks reproduce the coupling between agents for that prioritization.
        Teal edges run between blocks and connect the successive computations of one and the same agent, since an agent can work on only one prioritization at a time.
        A light gray source vertex S on the left has edges to all vertices of the first column, and a light gray sink vertex T on the right receives edges from all vertices of the last column, so that the networked computation time is the longest path from S to T.
    }
    \label{fig:evaluation:computation-graph}
\end{figure}

This additional relationship needs to be addressed in the computation graph as illustrated in \cref{fig:evaluation:computation-graph} for the coupling graphs of \cref{fig:method:coupling-graph}.
The edges in black indicate the couplings of agents in the three distinct prioritizations.
Since an agent can only compute one solution at a time, the subgraphs of these prioritizations are connected by additional edges in teal accounting for the sequential computation.
All edges in black and teal are weighted with the computation time $\tComp^{(i)}$ of an edge's respective starting vertex.
The weight of the longest path in this graph corresponds to the computation time of the \ac{ncs}.

\subsubsection{Computation Time Analysis}\label{sec:evaluation:time-planning}
\begin{figure}
    \centering
    \includegraphics{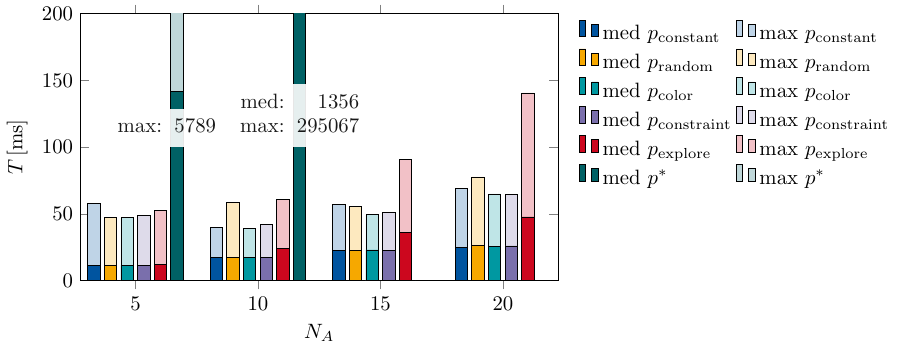}
    \caption{
        Comparison of the networked computation time among multiple prioritization algorithms.
        There is no value for $\fcnPrio^{*}$ for 15 and 20 vehicles as the optimization did not terminate.
    }
    \Description{
        Grouped bar chart.
        The horizontal axis is the number of agents with the four groups 5, 10, 15, and 20; the vertical axis is the networked computation time in milliseconds, cut off at 200 milliseconds.
        Each group holds twelve bars: the median and the maximum computation time for each of the six prioritization algorithms constant, random, color, constraint, explore, and optimal.
        The medians of constant, random, color, and constraint are nearly identical and grow from about 11 milliseconds at 5 agents to about 25 milliseconds at 20 agents.
        Our explorative prioritization is only slightly slower, from about 12 to about 48 milliseconds, while its maximum stays below about 141 milliseconds.
        The optimal prioritization is far above all others, with a median of about 142 milliseconds at 5 agents and about 1356 milliseconds at 10 agents, exceeding the axis range, and is missing for 15 and 20 agents because the optimization did not terminate.
    }
    \label{fig:evaluation:time}
\end{figure}

For five and ten vehicles, the maximum computation time of our approach is considerably lower than that of the optimal prioritization, while achieving a similar solution quality.
For a larger number of vehicles, planning with optimal prioritization is computationally too expensive, as the number of distinct prioritizations grows rapidly with the number of vehicles.
As shown in \cref{fig:evaluation:time}, our approach scales well with the number of vehicles and only slightly increases the computation time compared to the baseline prioritization.
In particular, the median computation time is only slightly increased.

\begin{figure}
    \centering
    \includegraphics{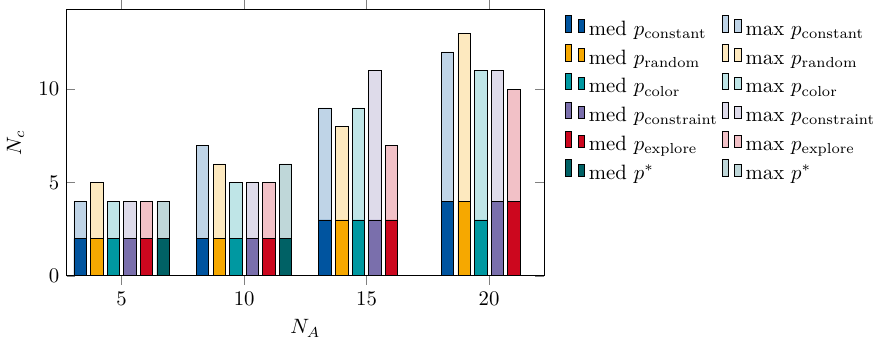}
    \caption{
        Comparison of the number of agent classes among multiple prioritization algorithms.
        The value of $\fcnPrio^{*}$ corresponds to the number of agent classes of the selected prioritization.
        There is no value for $\fcnPrio^{*}$ for 15 and 20 vehicles as the optimization did not terminate.
    }
    \Description{
        Grouped bar chart.
        The horizontal axis is the number of agents with the four groups 5, 10, 15, and 20; the vertical axis is the number of agent classes.
        Each group holds twelve bars: the median and the maximum number of agent classes for each of the six prioritization algorithms constant, random, color, constraint, explore, and optimal.
        The medians are almost identical across all algorithms and grow slowly with the number of agents, from 2 agent classes at 5 and 10 agents to 3 at 15 agents and 3 or 4 at 20 agents.
        The maxima are considerably larger and also grow with the number of agents, from about 4 or 5 at 5 agents to between 10 and 13 at 20 agents.
        The optimal prioritization is missing for 15 and 20 agents because the optimization did not terminate.
    }
    \label{fig:evaluation:levels}
\end{figure}

The anytime \ac{mcts} algorithm we use for planning has a fixed limit of 6000 expansions.
From the numerical experiment data, we get an average time per expansion of \qty{3}{\micro\second}, which results in an expected computation time of \qty{18}{\milli\second}.
If the computation times of all agents were exactly equal, the networked computation time in \cref{fig:evaluation:time} would scale with the number of agent classes in \cref{fig:evaluation:levels}.
Looking at the example of $\fcnPrio_{\text{constant}}$ with 20 vehicles, the maximum number of agent classes is 11, from which we expect a networked computation time of \qty{198}{\milli\second}.
The actual maximum networked computation time is only \qty{69}{\milli\second}.
This discrepancy is due to dissimilar computation times among agents.

Dissimilarities in the computation time are due to the fact that in some instances of planning, the search tree has been fully explored with fewer expansions than the limit of 6000.
A primary example of such a case is when a vehicle is blocked by others.
This example is particularly noticeable when the number of agent classes is high, as vehicles are in close proximity to one another.
Interestingly, this effect decreases the maximum networked computation time compared to the expected networked computation time from the number of agent classes for all single-shot prioritizations, but not for the proposed prioritization $\fcnPrio_{\text{explore}}$.
For the explanation of this phenomenon we need to look at the computation graph $\graphDirected^{\prime}$ displayed in \cref{fig:evaluation:computation-graph}.
Since we weight all outgoing edges with the computation time associated with solving the vertex's planning problem, the networked computation time is determined by the longest path in $\graphDirected^{\prime}$.
The edges in teal chain the $\numLevels$ computations of each agent, so the computations of a single agent already form a path in $\graphDirected^{\prime}$.
The weight of this path is the sum of that agent's computation times across all prioritizations.
It is therefore sufficient that a single agent requires the full expected computation time in each of its computations for the networked computation time to be near its expected value.

\begin{remark}
    If the computation time of all agents for each planning problem of every prioritization is exactly equal, there is no overhead in computation time for our approach.
\end{remark}

In its present form, our framework lays the foundations for guaranteed real-time planning.
Even though our approach finishes before the sample time $\tSample$ in our experiments, it does not account for the worst-case execution time that can occur when the number of agent classes equals the number of agents.
Two natural approaches for real-time guarantees suggest themselves.
One, limiting the number of agent classes to a desired value by enabling additional parallel computations while ensuring safety by restricting the solution space~\citep{scheffe2024limiting}.
Two, limiting the number of expansions and thus the expected computation time per agent depending on the number of agent classes.
Both approaches are only possible since coupling and prioritization are performed before planning in the framework used in this article.

\subsection{Experiment}\label{sec:evaluation:experiment}

\begin{figure}
    \centering
    \includegraphics[width=0.8\linewidth]{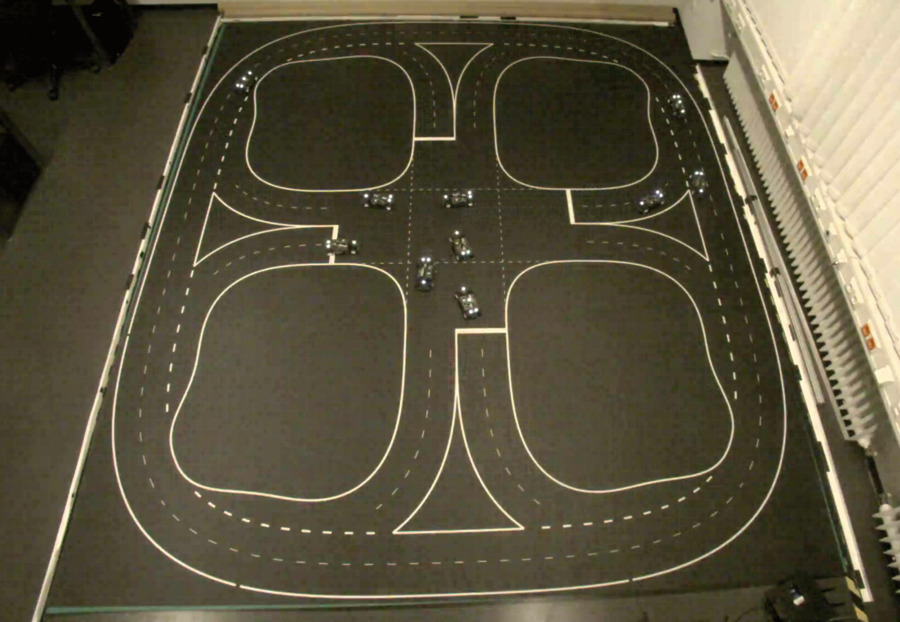}
    \caption{
        Experiment in the \ac{cpmlab} with ten vehicles.
    }
    \Description{
        Photograph taken from above of the road map of the Cyber-Physical Mobility Lab.
        A dark, roughly square surface is printed with a white road network consisting of an outer oval loop, four inner loops, and a crossing of two two-lane roads in the middle, with dashed white lane markings.
        Ten small model vehicles drive on the roads, most of them clustered in and around the central intersection.
    }
    \label{fig:evaluation:experiment}
\end{figure}

We conducted an experiment with ten vehicles in the \ac{cpmlab} \citep{kloock2021cyberphysical}.
The motion planning is executed on ten PCs connected via Ethernet, with each running the motion planning algorithm in MATLAB.
A snapshot of the experiment is shown in \cref{fig:evaluation:experiment}; a video is available online\footnote{\href{https://\videolink}{\videolink}}.
To achieve real-time capability, we combine our approach with our previous work of limiting the number of agent classes \citep{scheffe2024limiting}.
We chose a limit of $\numLevels = 2$.

In our experiment, we achieve a networked computation time with a median of \qty{72}{\milli\second} and a maximum of \qty{111}{\milli\second} across all time steps.
With our time step duration of $\tSample=\qty{200}{\milli\second}$, this networked computation time allows enough leeway for other subordinate computations and for communication.

Although communication time is excluded from the main algorithmic analysis, we report it here as an implementation-specific observation.
The communication time between the motion planners in the experiment has a median of \qty{5}{\milli\second} and a 99th percentile of \qty{25}{\milli\second}.
This communication time is surprisingly high considering the PCs are connected via Ethernet.
In a simulation study of three scenario variations with 20 PCs connected via Ethernet, we observed a median communication time of \qty{4}{\milli\second}, and a maximum of \qty{33}{\milli\second}.
With our simultaneous approach, the communication time increased to \qty{11}{\milli\second} in the median, and to \qty{157}{\milli\second} in the maximum.

These measurements are reported for completeness and to characterize the prototype implementation; they are not used to support the main comparison of this article, which focuses on planning performance under the scope assumptions stated in the introduction.

In our approach, the number of prioritizations explored is equal to the number of agent classes.
For each prioritization, communication is required.
Consequently, this results in an increase in the total communication effort by a factor of the number of agent classes.
However, the plans regarding each prioritization can be communicated in parallel, resulting in the same number of $\numLevels-1$ sequential communications.
Additionally, agents communicate the cost of their solution to determine the solution quality of each prioritization.
In practice, the exchanged data is small, as it only consists of the trajectory over the prediction horizon and the cost of the planning problem.
Our implementation with MATLAB and the ROS~2 toolbox is not intended for efficiency in communication and seems to struggle with this higher number of messages, resulting in high latencies.

\section{Conclusion}\label{sec:conclusion}
This section summarizes the contributions of this article and outlines directions for future work.

\subsection{Summary}

We presented an approach to utilize idle time in prioritized planning to simultaneously compute with multiple prioritizations.
With our approach, we can achieve high solution quality with significantly less computational effort than that of sequential computation of the considered prioritizations.
Additionally, our approach is general, as it does not rely on domain-specific heuristics.

Our approach can simultaneously examine as many prioritizations as there are agent classes.
The higher the number of agent classes, the more prioritizations are examined.
However, the number of distinct prioritizations increases more quickly with the number of agent classes.
Therefore, our approach is more likely to find the optimal prioritization with a lower number of agent classes.
Additionally, a lower number of agent classes results in a smaller increase in computation time.
These benefits make the approach well-suited to be combined with limiting the number of agent classes for real-time computation \citep{scheffe2024limiting}.

\subsection{Future Work}\label{sec:conclusion:future-work}
Our approach is general and can thus be transferred to prioritized computations in other applications, where communication is required and planning time outweighs communication time.
It needs to be analyzed how well the approach transfers, especially regarding the communication effort and the computation time.

Since the planning problems of multiple prioritizations are solved simultaneously, the communication effort compared to computing with a single prioritization is multiplied by the number of agent classes.
In our experiments, agents communicated via LAN.
It seems likely that the communication effort is decreased if computations run on the same machine, and increased if agents communicate via WLAN.
In any case, the gain in solution quality must outweigh the increase in communication effort compared to computing with a single prioritization.
Further studies are required to verify that communication can be time-efficient enough to warrant the distributed execution of our approach for a large number of agents.

We presented solving \iac{ocp} as a control method to find control inputs and to generate a motion plan.
However, our approach works well with any control method in which agents have similar computation times.
Therefore, the approach could also be evaluated with different control methods.

We presented simulations with up to 20 agents and eight agent classes, and an experiment with ten agents and two agent classes.
Conducting experiments with a larger number of agents and agent classes would be interesting to study the convergence to the optimal prioritization.
An application with a convex \ac{ocp} might be suitable to still achieve real-time capable computations.

Our algorithm to build a computation schedule matrix selects prioritizations randomly from all valid prioritizations.
The algorithm might improve if it instead selects prioritizations based on knowledge of the solution quality of prioritizations it has solved before.

We aimed to increase the solution quality without increasing the computation time by utilizing unused computation resources in distributed \ac{pp}.
Thus, we construct a Latin square to explore $\numLevels$ prioritizations in $\numLevels$ sequential computations.
However, the approach can be generalized to a desired number of sequential computations or a desired number of explored prioritizations $N_{\text{explored}}$, making the computation schedule matrix in a Latin rectangle.
Then, the number of explored prioritizations can be selected according to the available computation time.

Our approach for simultaneous computation of multiple prioritizations is intended for \ac{pp} where the coupling graph can be determined before planning.
With extensions to limit the number of agent classes \citep{scheffe2024limiting}, this kind of \ac{pp} can realize real-time control.
When constructing a coupling graph a priori is impossible, \ac{pp} with ad-hoc construction of a coupling graph can be used \citep[e.g.,][]{ma2019searching,chan2023greedy}.
Adapting our algorithm for this setting could increase the utilization of available computation resources.

\begin{acks}
    This research was funded by the Deutsche Forschungsgemeinschaft (DFG, German Research Foundation) within the Priority Program SPP 1835 ``Cooperative Interacting Automobiles'' (grant number: KO 1430/17-1).
\end{acks}

\printbibliography

\appendix
\section{Supplementary Materials}
\noindent
\begin{tabular}{@{} l l @{}}
    \textbf{Code}  & \href{https://github.com/pscheffe/p-dmpc}{github.com/pscheffe/p-dmpc} \\
    \textbf{Video} & \href{https://\videolink}{\videolink}                                 \\
\end{tabular}

\end{document}